\documentclass[structabstract]{aa}

\usepackage{amsmath}          
\usepackage{dcolumn}          
\usepackage{natbib}           
\usepackage{graphicx}         
\usepackage{txfonts}          
\usepackage{mathrsfs}         
\usepackage{rotating}         
\usepackage{subfigure}        
\usepackage{lscape}           
\usepackage{afterpage}        

\bibpunct{(}{)}{;}{a}{}{,}    
\newcolumntype{.}{D{.}{.}{-1}}
\newcolumntype{d}[1]{D{.}{.}{#1}}

\newcommand{\mcl}[3]{\multicolumn{#1}{#2}{#3}}
\newcommand{\mrm}[1]{\ensuremath{\mathrm{#1}}}
\newcommand{\nnhp}{\ensuremath{\mathrm{N}_2\mathrm{H}^+}}
\newcommand{\nndp}{\ensuremath{\mathrm{N}_2\mathrm{D}^+}}
\newcommand{\innhp}{\ensuremath{^{15}\mathrm{NNH}^+}}
\newcommand{\ninhp}{\ensuremath{\mathrm{N}^{15}\mathrm{NH}^+}}
\newcommand{\isot}[2]{\ensuremath{^{#1}\mathrm{#2}}}

\providecommand*{\mut}[1]{\ensuremath{\,\mathrm{#1}}}

\newcommand{\wsxj}[6]{\ensuremath{\left\{\begin{matrix} #1 & #2 & #3 \\ #4 & #5 & #6 \end{matrix}\right\}}}
\begin{document}

\title{Detection of $^{15}$NNH$^+$ in L1544: non-LTE modelling of dyazenilium
       hyperfine line emission and accurate $^{14}$N/$^{15}$N values%
       \thanks{Based on observations carried out with the IRAM 30\,m Telescope. 
               IRAM is supported by INSU/CNRS (France), MPG (Germany) and IGN (Spain).}}

\author{L.~Bizzocchi\inst{1} \and P.~Caselli\inst{2} \and E.~Leonardo\inst{1}
        \and L.~Dore\inst{3}}

\institute{Centro de Astronomia e Astrof\'isica, Observat\'orio Astron\'omico de Lisboa,
           Tapada da Ajuda, 1349-018 Lisboa (Portugal). 
           \email{[bizzocchi,elle]@oal.ul.pt}
           \and
           School of Physics and Astronomy, University of Leeds, Leeds LS2 9JT (UK). 
           \email{P.Caselli@leeds.ac.uk}
           \and
           Dipartimento di Chimica ``G.~Ciamician'',  Universit\`a di Bologna,
           via F.~Selmi 2, I-40126 Bologna (Italy). 
           \email{luca.dore@unibo.it}}

\titlerunning{\isot{14}{N}/\isot{15}{N} isotopic ratio in L1544}
\authorrunning{L. Bizzocchi}

\abstract
{Samples of pristine Solar System material found in meteorites and interplanetary 
 dust particles are highly enriched in \isot{15}{N}. 
 Conspicuous nitrogen isotopic anomalies have also been measured  in comets, and 
 the \isot{14}{N}/\isot{15}{N} abundance ratio of the Earth is itself larger than the 
 recognised pre-solar value by almost a factor of two.  
 Ion--molecules, low-temperature chemical reactions in the proto-solar nebula have 
 been repeatedly indicated as responsible for these \isot{15}{N}-enhancements.}
{We have searched for \isot{15}{N} variants of the \nnhp\ ion in L1544, a prototypical 
 starless cloud core which is one of the best candidate sources for detection owing 
 to its low central core temperature and high CO depletion.
 The goal is the evaluation of accurate and reliable \isot{14}{N}/\isot{15}{N} ratio 
 values for this species in the interstellar gas.}
{A deep integration of the \innhp$\:(1-0)$ line at 90.4\,GHz has been obtained with 
 the IRAM 30\,m telescope. 
 Non-LTE radiative transfer modelling has been performed on the $J = 1 - 0$ emissions 
 of the parent and \isot{15}{N}-containing dyazenilium ions, using a Bonnor--Ebert sphere 
 as a model for the source.}
{A high-quality fit of the \nnhp\:(1--0) hyperfine spectrum has allowed us to derive a 
 revised value of the \nnhp\ column density in L1544.
 Analysis of the observed \ninhp\ and \innhp\ spectra yielded an abundance ratio 
 $N(\ninhp)/N(\innhp) = 1.1\pm 0.3$\@.
 The obtained \isot{14}{N}/\isot{15}{N} isotopic ratio is $\sim 1000\pm200$, suggestive of a 
 sizeable \isot{15}{N} depletion in this molecular ion.
 Such a result is not consistent with the prediction of present nitrogen chemical models.}
{As chemical models predict large \isot{15}{N} fractionation of \nnhp, we suggest that 
 \isot{15}{N}\isot{14}{N}, or \isot{15}{N} in some other molecular form, is preferentially 
 depleted onto dust grains.}


\maketitle

\section{Introduction} \label{sec:intro}
\indent\indent
Determination of the abundance ratios of the stable isotopes of the light elements 
in different objects of the Solar System is one of the key elements to understand 
its origin and early history \citep[see, e.g.][]{Caselli-AAr12-Astroch}.
Nitrogen, the fifth or sixth most abundant element in the Sun 
\citep[after H, He, C, O, and maybe Ne,][]{Asplund-ARAA09-Sun}, is particularly 
intriguing because its isotopic composition shows large variations whose interpretation 
is still controversial \citep{Aleon-APJ10-Nisot,Adande-ApJ12-N15}.

Recent laboratory analysis of the solar wind particles collected by the \emph{Genesis} 
spacecraft \citep{Marty-Sci11-N15} yielded a \isot{14}{N}/\isot{15}{N} ratio of $441\pm 6$,
higher than in any Solar System object, but probably representative of the proto-solar 
nebula value, being comparable to that measured in Jupiter's atmosphere 
\citep[$435\pm57$,][]{Owen-ApJ01-N,Fouch-Ica04-NJup} 
and in osbornite (TiN) calcium-aluminium-rich inclusion from the CH/CB chondrite 
\emph{Isheyevo} \citep[$424\pm3$,][]{Meibom-ApJ07-Chondr}.
The \isot{14}{N}/\isot{15}{N} value of the terrestrial atmosphere is significantly lower 
($\sim 272$), and larger \isot{15}{N} enrichments were measured in cometary nitrile-bearing 
molecules \citep{Arpi-Sci03-Ncomets,Bock-ApJ08-17PHolm,Manfroid-AA09-CN} and in primitive 
chondritic materials \citep[up to 50, e.g.,][]{Briani-LPI09-n15,Bonal-GCA10-Ishe}.

A common but still debated interpretation considers such variations as an inheritance 
of the proto-solar chemistry: low-temperature ion--molecule reactions 
\citep[e.g.,][]{Millar-PlSS02-Dchem} in the interstellar medium (ISM) were proven to 
cause large isotopic excesses for D in organic molecules, and have repeatedly been 
proposed as the cause of the observed \isot{15}{N} excesses too.
However, in molecular clouds, gas-phase chemistry continually cycles nitrogen between 
atomic and molecular forms, equating the composition of the isotopic reservoirs.
Indeed, classical ion--molecule reaction models fail to predict major \isot{15}{N} 
enrichments \citep{Terz-MNRAS00-Nfrac}.
Specific conditions, such as strong and selective CO freeze-out 
\citep{Charn-ApJ02-Nfrac,Rodg-MNRAS08-Nfrac}, might overcome this difficulty and produce 
a ``nitrogen super-fractionation'' in cold ISM, capable in principle to account for the 
largest measured enhancements.

A way to assess the isotopic composition of the pre-solar gas is the measurements of the 
\isot{14}{N}/\isot{15}{N} ratio in other proto-stellar systems.
Recent observations of \isot{15}{N}-bearing molecules found no significant fractionation 
in NH$_3$ ($\isot{14}{N}/\isot{15}{N} = 350-500$, \citealt{Gerin-ApJ09-15NH2D}; 
$334\pm 50$, \citealt{Lis-ApJ10-NH3}) toward pre-stellar cores and proto-stellar
envelopes, and in \nnhp\ \citep[$\isot{14}{N}/\isot{15}{N} = 446\pm 71$,][]{Bizz-AA10-L1544}
toward the prototypical starless cloud core L1544. 
Conversely, various preliminary measurements on nitrile species towards pre-stellar cores 
showed that they are highly enriched in \isot{15}{N} (between 70 and 380, 
\citealt{Milam-LPI12-Nfract}; \citealt{HilBl-Ica13-15N}; Bizzocchi unpublished).
Also, similar values have been found by \citet{Adande-ApJ12-N15} in HNC observations of 
massive star-forming regions across the Galaxy.

Another problem to directly link pre-stellar core chemistry and the Solar System composition 
comes from the poor correlation between D- and \isot{15}{N}-enhancements observed in some 
pristine materials \citep{Busem-MPSS06-Chondr,Robert-MPSS06-Chondr}, whereas it is clear 
that the chemical processes invoked to account for nitrogen fractionation should also produce 
enormous deuterium enhancements.

\citet{Wirst-ApJ12-Nspin} considered the spin-state dependence in ion--molecule reactions 
involving the \emph{ortho} and \emph{para} forms of H$_2$ and succeeded in reproducing the 
differential \isot{15}{N}-fractionation observed in amine- and nitrile-bearing compounds, 
as well as the overall lack of correlation with hydrogen isotopic anomalies.
The authors pointed out that, in cold interstellar environments, the 
\emph{ortho}-to-\emph{para} ratio of H$_2$ plays a pivotal role in producing a diverse 
range of D--\isot{15}{N} fractionation in precursors molecules, thus providing a strong 
support for the astrochemical origin of nitrogen isotopic anomalies.
However, this hypothesis still requires a sound verification since so far, observations 
of \isot{15}{N}-isotopologues in the ISM are rather sparse.
An extended survey of the \isot{14}{N}/\isot{15}{N} ratio targeting starless clouds in 
different evolutionary phases, and possibly in different environmental conditions, is
thus desirable.

Obtaining accurate determinations of the nitrogen isotopic ratio in the ISM is 
problematic: \isot{15}{N}-bearing species produce typically very weak emissions, 
which requires time consuming high sensitivity observations. 
Moreover, the rotational spectra of common N-containing species are usually optically 
thick and the line intensities are not reliable indicators of the molecular abundance.
For nitrile molecules, this latter difficulty may be overcome using less abundant 
\isot{13}{C} variants as proxies for the parent species and then deriving the 
\isot{14}{N}/\isot{15}{N} ratio from an assumption for the \isot{12}{C}/\isot{13}{C} 
\citep[e.g.,][]{Dahmen-AA95-Nratio}.
Another route is to use the hyperfine structure analysis to evaluate the optical depth 
of the parent species, thus allowing for a more direct determination of the nitrogen 
isotope ratio \citep{Savage-ApJ02-Cratio,Adande-ApJ12-N15}. 

In a previous letter, \citet{Bizz-AA10-L1544} reported the detection of \ninhp\ in L1544.
The analysis was carried out assuming local thermodynamic equilibrium (LTE) conditions
and yielded a \isot{14}{N}/\isot{15}{N} ratio of $446\pm 71$.
This is indicating the absence of nitrogen fractionation in the dyazenilium ion, a result
not consistent with the prediction of the chemical model of \citet{Gerin-ApJ09-15NH2D}
and \citet{Wirst-ApJ12-Nspin}.

The chosen target, L1544, is a prototypical starless core on the verge of the star-formation 
\citep{WT-MNRAS99-SFIII,Caselli-ApJ02-L1544k}.
Its density structure, low central temperature, and high CO depletion make it an ideal 
laboratory to study the isotopic fractionation processes.
Also, an accurate model for its internal structure and dynamics has been proposed by 
\citet{Keto-MNRAS10-SC} and then successfully used to analyse the H$_2$O emission 
observed by \emph{Herschel} \citep{Caselli-ApJ12-H2O}.
In this paper we present the detection of \innhp\ isotopologue in L1544 together with a 
full non-LTE radiative transfer treatment of the dyazenilium ions aimed at the evaluation 
of accurate and reliable values of the \isot{14}{N}/\isot{15}{N} ratio in this species.

The paper is organised as follows: in \S~\ref{sec:obs} we describe the technique used for
observations, and in \S~\ref{sec:results} we summarise our direct observational results.
\S~\ref{sec:model} is devoted to the description of the Monte Carlo radiative transfer 
modelling of \nnhp\ and its \isot{15}{N}-variants, and the derivation of their molecular 
abundances.
In \S~\ref{sec:disc} we discuss the implications for the chemical models and in 
\S~\ref{sec:conc} we summarise our conclusions.

\section{Observations} \label{sec:obs}
\indent\indent
The observations towards L1544 were carried out with the IRAM 30\,m antenna, located at 
Pico~Veleta (Spain) during observing sessions in June~2009 and July~2010.
The $J = 1 - 0$ transition of \innhp\ was observed with the EMIR receiver in the E090 
configuration tuned at 90\,263.8360\,MHz and using the lower-inner side-band.
The hyperfine-free rest frequencies were taken from the most recent laboratory investigation
of \isot{15}{N}-dyazenilium species \citep{Bizz-AA09-N2H+}.
Scans were performed in frequency switching mode, with a throw of $\pm$7\,MHz; the backend 
used was the VESPA correlator set to a spectral resolution of 20\,kHz (corresponding to 
0.065\,km\,s$^{-1}$) and spectral bandpass of 20\,MHz.
We tracked the L1544 continuum dust emission peak at 1.3\,mm, where we previously detected 
the other \isot{15}{N}-containing isotopologue, \ninhp\ \citep{Bizz-AA10-L1544}.
The J2000 coordinates are: $\mrm{RA} = 05^\mrm{h}04^\mrm{m}17.21^\mrm{s}$,  
$\mrm{Dec} = 25^\circ 10'42.8''$ \citep{Caselli-ApJ02-L1544k}.
The telescope pointing was checked every two hours on nearby bright radio quasars and was 
found accurate to 3--4\arcsec; the half power beam width (HPBW) at the line frequency is
27\arcsec.

In may~2009 we spent 4.25\,hours on source with average atmospheric condition 
($\tau\sim 0.1$), while during the summer 2010 session we observed for further 23.9\,hours 
with good weather conditions ($\tau < 0.05$); all those scans were summed together for a 
total of 28.15~hours of on-source telescope time.
Horizontal and vertical polarizations were simultaneously observed and averaged together 
to produce the final spectrum, which was then rescaled in units of $T_\mrm{mb}$ assuming 
a source filling factor of unity and using the forward and main beam efficiencies 
appropriate for $91\mut{GHz}$: $F_\mrm{eff}=0.95$ and $B_\mrm{eff}=0.75$, respectively.
The rms noise level achieved was about 3\,mK, close to that obtained for the $J=1-0$ 
line of the species \ninhp\ toward the same line of sight \citep{Bizz-AA10-L1544}.

The same backend configuration was also employed to collect new data for the $J = 1 - 0$ 
transition of the main isotopologue, \nnhp, with the EMIR receiver tuned at 93\,173.4013\,MHz\@.
This line was observed shortly at the beginning of each telescope session for a total 
integration time of $\sim$56\,min\@. 
The final rms noise level is 15\,mK, resulting in a high signal-to-noise spectrum, well suited
for modelling purposes.

%
\begin{table*}[tbh]
  \caption[]{Predicted hyperfine frequencies, estimated $1\sigma$ uncertainties, and relative 
             line intensities for the $J = 1 - 0$ transition of \ninhp\ and \innhp\ 
             \citep{Bizz-AA09-N2H+}.}
  \label{tab:hfs-spec}
  \centering 
  \begin{tabular}{c c . . .}
    \hline\hline \noalign{\smallskip}
    Isotopologue & 
    $F' - F$     & 
    \mcl{1}{c}{Frequency} &
    \mcl{1}{c}{Uncertainty} &
    \mcl{1}{c}{Relative intensity} \\
                 & 
                 &
    \mcl{1}{c}{(MHz)}  &
    \mcl{1}{c}{(kHz)}  &
                       \\[0.5ex]
    \hline \noalign{\smallskip}
    \ninhp & 1 - 1 &  91\,204.2602   &  0.9  &  1.000 \\
           & 2 - 1 &  91\,205.9908   &  0.8  &  1.667 \\
           & 0 - 1 &  91\,208.5162   &  1.2  &  0.333 \\[0.5ex]
    \innhp & 1 - 1 &  90\,263.4870   &  0.9  &  1.000 \\
           & 2 - 1 &  90\,263.9120   &  0.6  &  1.667 \\
           & 0 - 1 &  90\,264.4972   &  1.6  &  0.333 \\
    \hline 
  \end{tabular}
\end{table*}
%
\begin{table*}[tbh]
  \caption[]{Results of the CLASS HFS fit on the observed spectral profile of the
             \isot{15}{N}-bearing dyazenilium isotopologues observed towards L1544.
             Numbers in parentheses refer to $1\sigma$ uncertainties in units of the 
             last quoted digit.}
  \label{tab:hfs-mult}
  \centering 
  \begin{tabular}{c . d{7} d{5} . c .}
    \hline\hline \noalign{\smallskip}
    Line   & 
    \mcl{1}{c}{Rest frequency$^a$}     & 
    \mcl{1}{c}{$A$ coefficient$^b$}    &
    \mcl{1}{c}{$v_{LSR}$}              &  
    \mcl{1}{c}{$\int T_\mrm{mb}\mrm{d}v$\,$^c$}  & 
    \mcl{1}{c}{$\Delta v$\,$^d$}       \\
           & 
    \mcl{1}{c}{(MHz)}                  & 
    \mcl{1}{c}{$10^5\,$s$^{-1}$}       & 
    \mcl{1}{c}{(km$\,$s$^{-1}$)}       &  
    \mcl{1}{c}{(mK$\,$km$\,$s$^{-1}$)} &
    \mcl{1}{c}{(km$\,$s$^{-1}$)}       \\[0.5ex]
    \hline \noalign{\smallskip}
    \ninhp\,(1--0) & 91\,205.6945 & 3.19(32) &  7.299(14) &  21.4(22) & 0.507(43) \\
    \innhp\,(1-0)  & 90\,263.8354 & 3.42(35) &  7.203(10) &  18.3(13) & 0.409(21) \\
    \hline 
  \end{tabular}
  \begin{list}{}{}
    \item[$^a$] From \citet{Bizz-AA09-N2H+}.
    \item[$^b$] Calculated through Eq.~(5) of \citet{Bizz-AA09-N2H+} and using the data of the same paper.
    \item[$^c$] Integrated intensity summed over all the components.
    \item[$^b$] Gaussian FWHM. Assumed equal for all the components. 
  \end{list}
\end{table*}
%
\begin{figure*}[tbh]
 \centering
 \includegraphics[angle=0,width=17.5cm]{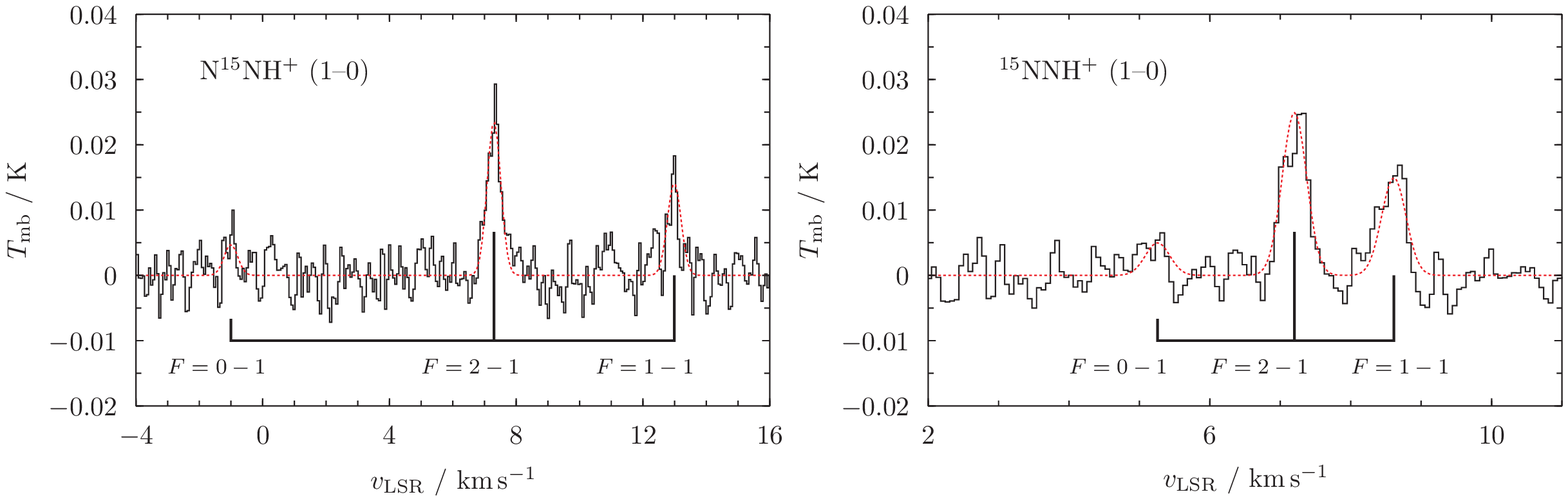}
 \caption{Spectra of the \isot{15}{N}-containing dyazenilium isotopologues observed towards
          L1544 (black histogram), and computed spectral profiles resulting from the HFS 
          fits (red dashed curves). 
          Superimposed black lines indicate the position and relative intensity of the 
          hyperfine components.
          Left panel: \ninhp\:(1--0) transition (reproduced from \citealt{Bizz-AA10-L1544}).
          Right panel: \innhp\:(1--0) transition, observed in July 2010 at IRAM\:30m\@. 
          The rms noise level is 2.6\,mK.}
 \label{fig:obs-spectra}
\end{figure*}

\section{Results} \label{sec:results}
\indent\indent
The two \isot{15}{N}-dyazenilium spectra observed in L1544 are presented in 
Figure~\ref{fig:obs-spectra}.
Due to the smaller magnitude of the electric quadrupole coupling constant or the 
inner \isot{14}{N} atom, the hyperfine triplet of the \innhp\:(1--0) transition is 
spread over $\sim 1$\,MHz, much less than the other \isot{15}{N}-isotopologue.
For the sake of completeness, the frequencies and relative intensities of the $J = 1 - 0$ 
hyperfine transition of both \isot{15}{N}-bearing dyazenilium variants are reported 
in Table~\ref{tab:hfs-spec} \citep{Bizz-AA09-N2H+}.

The data were processed using the GILDAS\footnote%
{See GILDAS home page at the URL:\\ \texttt{http://www.iram.fr/IRAMFR/GILDAS}.} 
software \citep{Pety-SF05-GILDAS}.
After polynomial baseline subtraction, average line parameters were estimated by fitting 
Gaussian line profiles to the detected components using the HFS routine implemented in CLASS.
The hyperfine splittings and relative intensities of the $J = 1-0$ transitions of both 
\isot{15}{N} isotopologues were taken from Table~\ref{tab:hfs-spec} and kept fixed in the 
least-squares procedure.
Since the lines are optically thin within the HFS fitting errors, we forced the optical depth 
to have the value of~0.1, the minimum allowed in CLASS. 
This method has been adopted in the past \citep[e.g.,][]{Caselli-ApJ02-L1544i} to avoid highly 
uncertain values of the optical depth to affect the intrinsic line width, as the error on 
$\tau$ is not propagated in the evaluation of the line width. 

The derived parameters for \innhp\:(1--0) are gathered in Table~\ref{tab:hfs-mult}, where 
the results obtained from the previous \ninhp\:(1--0) observations are also reported for 
completeness.
From these data one can derive a systemic velocity $V_\mrm{LSR} = 7.203\pm 0.010$\,km\,s$^{-1}$
for \innhp\:(1--0) which compares well with the value derived previously for the \ninhp\:(1--0),
i.e.,  $V_\mrm{LSR} = 7.299\pm 0.014$\,km\,s$^{-1}$ \citep{Bizz-AA10-L1544}.    
The small discrepancies observed in source velocity and in the line width between the
\isot{15}{N}-species (these quantities agrees within 5$\sigma$), are likely to be attributed
to the low signal-to-noise ratio and the spectral resolution (0.067\,km\,s$^{-1}$).

\section{L1544 modelling and radiative transfer} \label{sec:model}
\indent\indent
In the following subsections we describe the radiative transfer treatment carried
out on L1544 to model the $J = 1-0$ emission of \nnhp, \ninhp, and \innhp, and to derive
accurate molecular abundances and $\isot{14}{N}/\isot{15}{N}$ ratios.
The physical model is based on that proposed by \citet{Keto-MNRAS10-SC} and updated with 
the inclusion of oxygen to interpret the far infrared emission of the water vapour in the 
same source \citep{Caselli-ApJ12-H2O}.    
Briefly, the core is described as a gravitationally contracting Bonnor--Ebert sphere;
the model includes radiative equilibrium from dust, gas cooling \emph{via} both molecular 
emission lines and grain collisional coupling, as well as a simplified molecular chemistry
\citep{Keto-ApJ08-SC}.

\begin{figure}[tbh]
  \centering
  \includegraphics[angle=0,width=9cm]{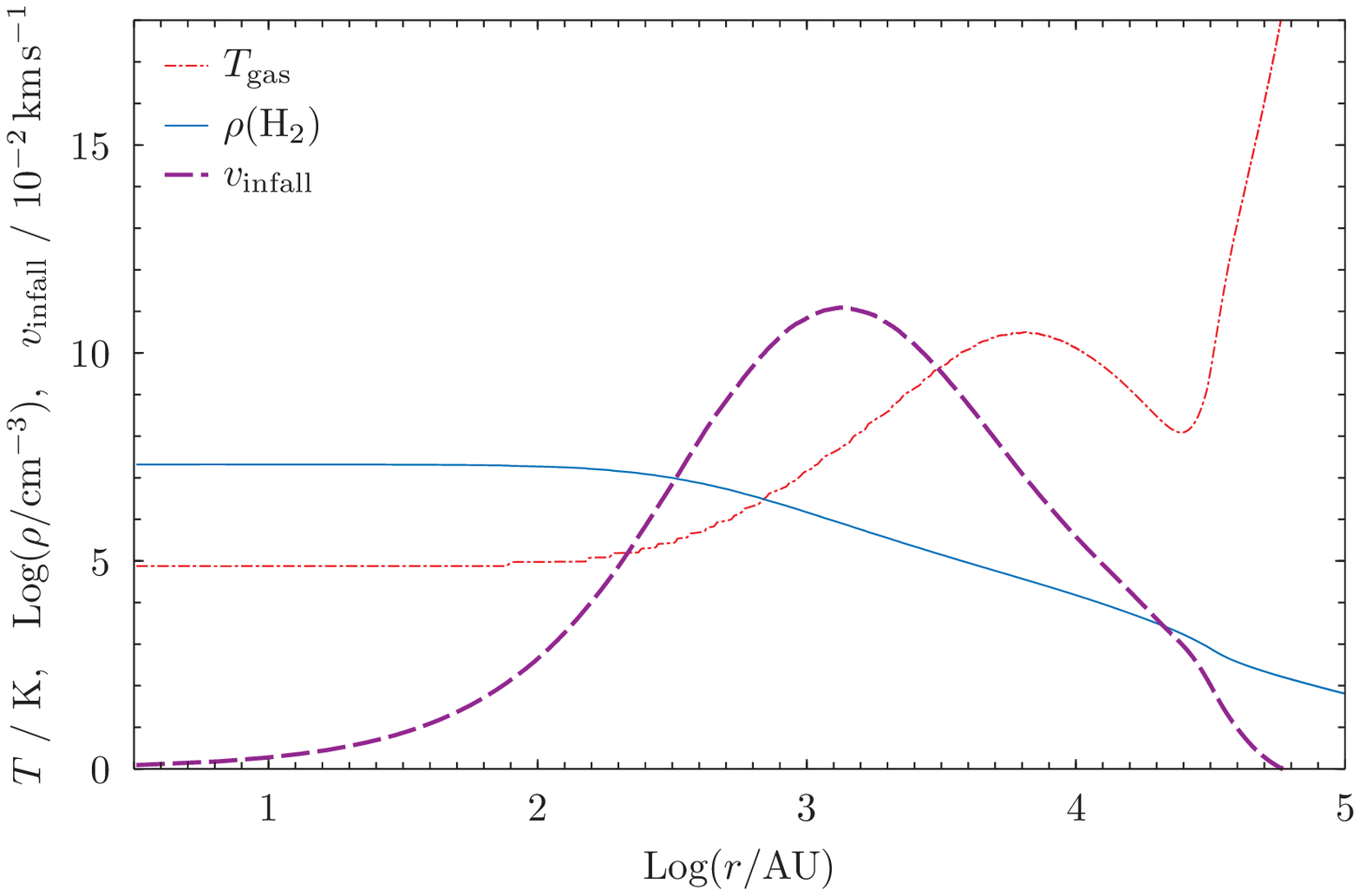}
  \caption{Temperature, density, and inward velocity profiles of L1544 used as model
           in the present radiative transfer modelling.
           The red dotted curve indicates the gas temperature in K, the blue curve is the log 
           of density in cm$^{-3}$, and the purple dashed curve represents the inward velocity
           in units of 0.01\,km\,s$^{-1}$.}
  \label{fig:model1}
\end{figure}

First, we used the original model with a central density of $2\times 10^7$\,cm$^{-3}$: 
its temperature, density, and inward velocity profile are shown in Figure~\ref{fig:model1}.
The H$_2$ column density, averaged over the 30\,m telescope main beam FWHM at 3\,mm, 
is $6.57\times 10^{22}$\,cm$^{-2}$. 
This value compares well with that derived by \citet{Crapsi-ApJ05-L1544} through dust 
continuum emission observation at 1.2\,mm; once averaged over the same beam (11\arcsec), 
our model yields $11.5\times 10^{22}$\,cm$^{-2}$, consistent with the observed value of 
$(9.4\pm 1.6)\times 10^{22}$\,cm$^{-2}$.

The radiative transfer calculations have been performed using the non-local thermodynamic
equilibrium (non-LTE) numerical code MOLLIE \citep{Keto-ApJ90-RT,Keto-APJ04-RT}.
We used here the updated version of the algorithm, able to treat in a proper way the 
issue of overlapping lines, thus allowing to better reproduce the non-LTE hyperfine ratios 
(excitation anomalies) observed in the \nnhp\ spectra of L1544 and in a number of other 
starless cores \citep{Caselli-ApJ95-N2H+,Daniel-ApJ07-N2H+II}.
For the present modelling we mostly relied on the hyperfine rates calculations of 
\citet{Daniel-JCP04-N2H+,Daniel-MNRAS05-N2H+} for \nnhp/He collisions; appropriate rate 
coefficients for H$_2$ partner were then obtained using a suitable scaling relation 
(see Appendix~\ref{sec:HypRates-gen}).

The cloud structure of L1544 was modelled with 3 nested grids, each composed of 48~cells.
The cell linear dimensions of each nested level are decreasing, i.e., the level~1 covers 
the whole source out to a radius of 66\,000\,AU, whereas the finer level~3 maps the inner 
16\,000\,AU of the core.
In each cell, the temperature, density, and gas kinematic parameters were taken from 
the model shown in Figure~\ref{fig:model1} and assumed constant.
A constant turbulent FWHM line width of 0.13\,km\,s$^{-1}$ 
\citep[as found by][]{Tafalla-ApJ02-SCmol} was added in quadrature to the thermal 
line width calculated in each model cell.

\subsection{N$_2$H$^+$\:(1--0)} \label{sec:n2h+}
\indent\indent
Radiative transfer modelling of \nnhp\ is not an easy task. 
Because of the two \isot{14}{N} nuclei, each rotational level is split by quadrupole 
interactions into nine sub-levels and the rotational transitions exhibit a complex 
structure of hyperfine components (i.e., 15 for $J=1-0$, 38 for $J=2-1$, etc.) with
various degrees of overlap between them.
The presence of the hyperfine structure (HFS) complicates the modelling. 
The relative populations of the hyperfine sub-levels may depart from their statistical
weights producing non-LTE intensity ratios \citep[e.g.,][]{Caselli-ApJ95-N2H+}.
Indeed, collisional coefficients for each individual hyperfine component typically have 
different values \citep[e.g.,][]{Buffa-MNRAS12-DCO+}, thus producing hyperfine selective 
collisional excitation \citep{Winnew-AA85-NH3}.
A second effect is the ``hyperfine line-trapping'', i.e., various components may have 
different optical depth, thus getting different amounts of radiative excitation.
The trapping becomes important at high optical depths and is accentuated by the overlap 
of the various hyperfine components.
\begin{figure}[tbh]
  \centering
  \includegraphics[angle=0,width=9cm]{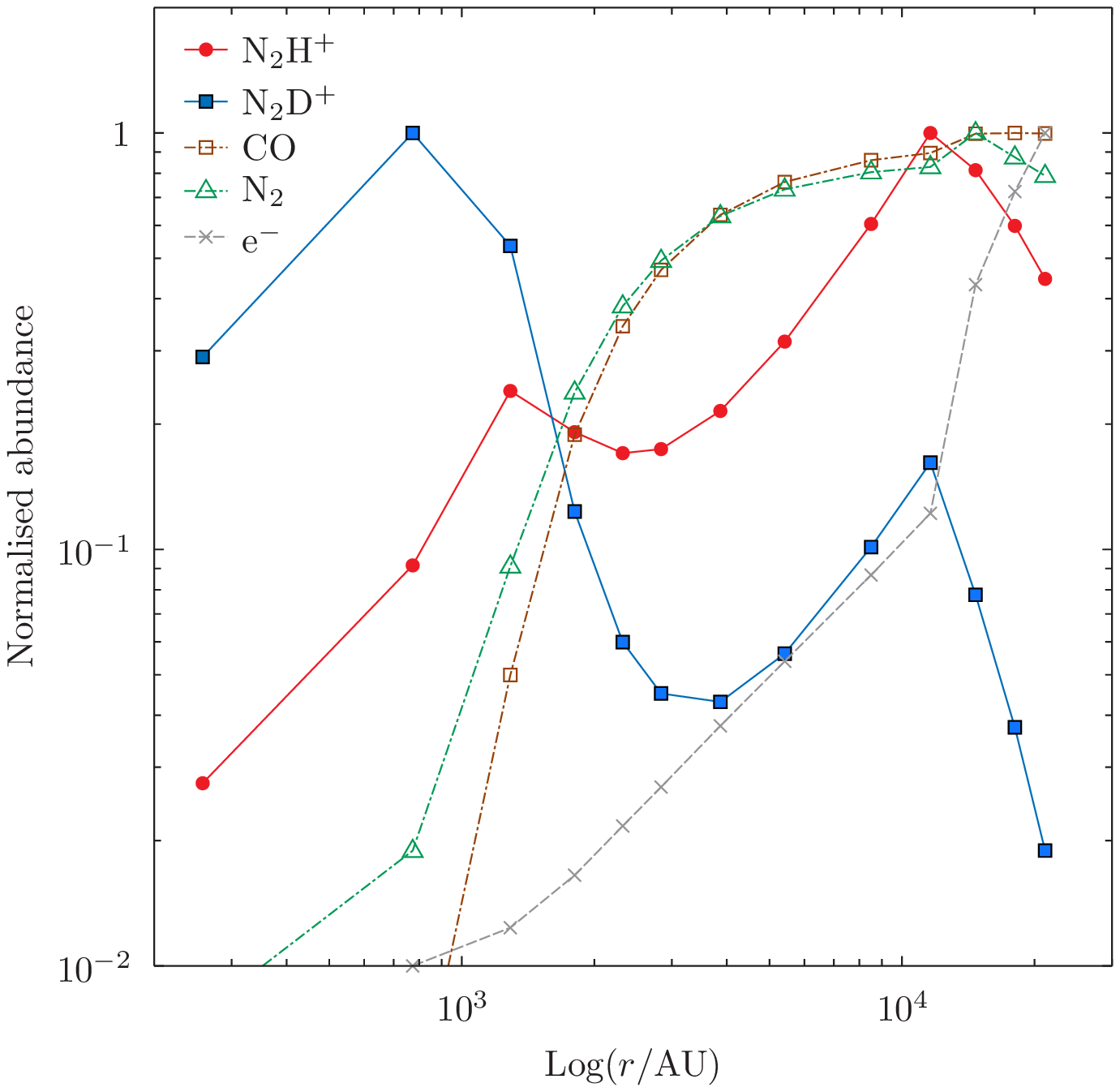}
  \caption{
           Abundance trends (normalised to the maximum values) predicted by the 
           \citet{Aikawa-IAU11-chem} models for dyazenilium and chemically related species.
           The \nnhp\ abundance decreases toward the cloud centre following the N$_2$ 
           freeze-out (green curve) onto dust mantels at high gas density.
           The concomitant steeper drops of free electron (grey curve) and CO (brown curve)
           give rise to the \nnhp\ abundance maxima at radii of $\sim 10^4$ and $\sim 10^3$ 
           astronomical units.
           \nndp\ shows a similar behaviour (blue curve) but it features a much stronger inner 
           peak owing to the large deuterium fractionation existing in the highly--CO depleted 
           central region.}
  \label{fig:abund}
\end{figure}

A Monte Carlo modelling of the \nnhp\:(1--0) emission line in L1544 and other 
starless cores has previously been performed by \citet{Tafalla-ApJ02-SCmol}.
They adopted a simplified treatment in which the above mentioned effects are ignored:
all the hyperfine sub-levels were assumed to be populated according to their statistical 
weights (no hyperfine selective collisional excitation was considered), and the hyperfine 
line trapping was neglected.
To justify such a simple approach, it was argued that line trapping is only significant 
at very large optical depths and, in any case, non-LTE intensity effects involved typically 
only 10--15\% of the emerging flux.
For most sources this modelling provided \nnhp\ spectra in good agreement with observations 
but it yielded a less satisfactory result for L1544.
Notably, the predicted spectrum was unable to reproduce the non-Gaussian line shape of the 
hyperfine lines which was attributed to the presence of two velocity components in the 
core \citep{Tafalla-ApJ98-L1544}. 
They were thus roughly modelled using a broader profile.

Another issue involves the \nnhp\ abundance profile.
So far, the source model used in radiative transfer studies of L1544 assumed a constant 
\nnhp\ abundance throughout the source 
\citep{Tafalla-ApJ98-L1544,Tafalla-ApJ02-SCmol,Keto-APJ10-HypRT}.
This hypothesis is reasonable, given the similarity between the observed L1544 dust 
continuum and \nnhp\:(1--0) emission maps \citep[see, e.g.][]{Tafalla-ApJ02-SCmol}.
Also, the observed radial profiles of the integrated line emission intensity
\citep[see Fig.~2 of][]{Caselli-ApJ99-COdep}, show that the abundance behaviour of \nnhp\ 
is markedly different from that of CO, which is known to suffer considerable depletion at
the high gas densities toward the core centre.
However, \citet{Caselli-ApJ02-L1544i} suggested that a certain amount of nitrogen depletion 
is also expected in the dense gas, thus the dyazenilium abundance is also likely to show a 
drop toward the L1544 centre \citep[see also][]{Bergin-ApJ02-CDCdep}.

Given the uncertainties in the spatial distribution of \nnhp\ (due to the poor spatial 
resolution), we investigated both constant and central-drop abundance profiles. 
In either instances, we ran a grid of models with varying standard \nnhp\ abundance, and 
sought the best fit of the observed line profiles.
\begin{figure}[tbh]
  \centering
  \includegraphics[angle=0,width=9cm]{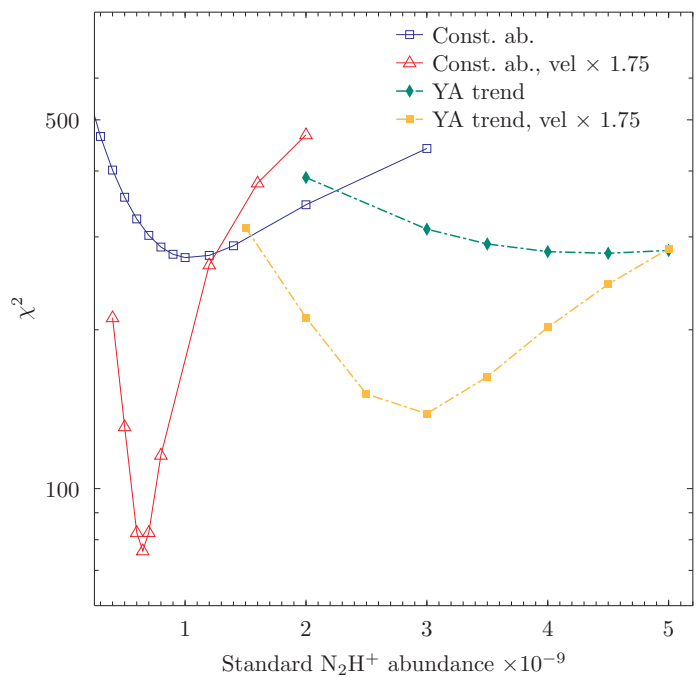}
  \caption{Weighted $\chi^2$ of the \nnhp\:(1--0) modellings using different standard 
          dyazenilium abundances and radial profiles.
          Blue curve: constant \nnhp\ abundance with respect to H$_2$;
          red curve: constant abundance and modified cloud infall velocity (see text).
          Green curve: radial \nnhp\ abundance profile shown in Figure~\ref{fig:abund};
          yellow curve: radial profile with modified cloud infall velocity.}
  \label{fig:chisq}
\end{figure}
\begin{figure*}[t]
  \centering
  \includegraphics[angle=0,width=16cm]{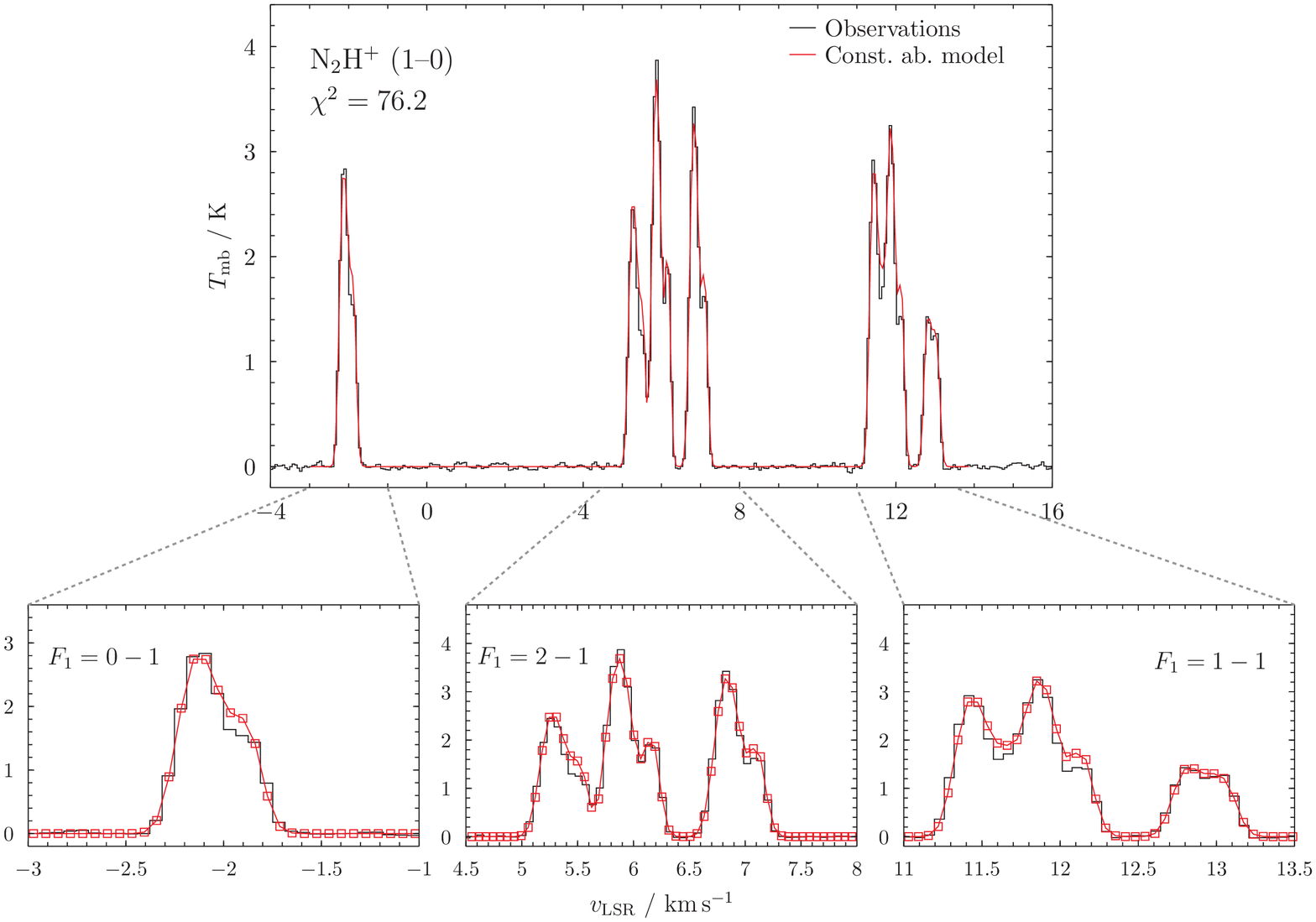}
  \caption{Observed \nnhp\:(1--0) spectrum in L1544 modelled with non-LTE code MOLLIE.
           The histograms show our observation data.
           The red curve shows the model spectrum computed using the ``altered'' dark 
           cloud model and \nnhp\ constant abundance throughout the core (see text).} 
  \label{fig:n2h+_10_best}
\end{figure*}

In the first set of models we used a constant dyazenilium abundance with respect to H$_2$, 
then we tested the \nnhp\ radial abundance profile calculated using the 
hydro-dynamical--chemical model of \citet{Aikawa-IAU11-chem}. 
The employed chemical network is the same as in \citet{Aikawa-ApJ12-Dchem}, but not the 
dynamics. 
In this paper, we assume that the physical structure is fixed at the model shown in 
Figure~\ref{fig:model1} and the chemistry is run until the C$^{18}$O column density reaches 
the observed value within the IRAM\,30m beam.
The predicted trends for a selection of chemical species are illustrated in 
Figure~\ref{fig:abund}. 
Dyazenilium is formed by proton transfer to N$_2$ and destructed by reaction with CO and 
recombination with electron.
As expected, the nitrogen freeze-out at smaller radii produces a overall decreasing \nnhp\ 
abundance trend, while the secondary peak at $\sim 1000$\,AU is due to the concomitant 
faster drop of the main depleting reactants. 

The ``goodness'' of each modelling was estimated using the quantity:
\begin{equation} \label{eq:chi2}
 \chi^2 = \sum_i^{N}\left\{\frac{1}{\sigma_\mrm{obs}}
                           \left[T_\mrm{mb,obs}(i) - T_\mrm{mb,model}(i)\right]
                    \right\}^2 \,,
\end{equation}
where $T_\mrm{mb}(i)$ are the brightness temperature of the observed and modelled 
spectra at the $i$-\emph{th} velocity channel and $\sigma_\mrm{obs}$ is the actual 
rms noise of the observation which is estimated to be 0.015\,K.
The sum in Eq.~\eqref{eq:chi2} run over all the velocity channels calculated by MOLLIE, 
covering an interval of 2\,km\,s$^{-1}$ around each hyperfine component.

Figure~\ref{fig:chisq} plots the resulting $\chi^2$ for varying standard \nnhp\ 
abundance and different abundance profiles.
The blue and green curves represent the $\chi^2$ trends obtained adopting constant \nnhp\ 
abundance, and the radial \nnhp\ abundance profile shown in Figure~\ref{fig:abund},
respectively.
The two models yield fits of comparable but not satisfactory quality, as both fail 
in reproducing the observed brightness of the thinner $F_1,F = (1,0)-(1,1)$ component.
Another weakness of these simulations is the overall poor agreement of the spectral profile, 
i.e., the calculated lines are systematically narrower than the observed ones.
This suggests that the inward velocity profile adopted as a model 
(see Figure~\ref{fig:model1}) could be underestimated.
In fact, the hydrodynamic calculations of \citet{Keto-MNRAS10-SC} result in idealised 
models of contracting Bonnor--Ebert spheres in quasi-static equilibrium and the contraction 
velocity was derived from noisier data. 

We have thus altered L1544 cloud model by increasing its inward velocity profile by a 
constant factor, and sought again the minimum $\chi^2$ by varying the \nnhp\ standard 
abundance.
Both constant and centrally decreasing \nnhp\ abundance profiles were tested.
Best fits were obtained using a velocity profile scaled with a constant 1.75 factor; the 
resulting trends are illustrated by the red and yellow curves in Figure~\ref{fig:chisq}.
A significant improvement was achieved for both profiles; the model with constant abundance 
drops to a narrow $\chi^2$ minimum of 76.2 and supersedes the best fit obtained with the 
centrally decreasing profile, which is almost twice poorer ($\chi^2\sim 140$).
The obtained best $\chi^2$ for the two models correspond to weighted rms of~0.54 and~0.98, 
indicating that in both cases the observed profile is reproduced within the average spectral 
noise.

We regard the spectrum obtained with the constant \nnhp\ abundance profile and a standard
abundance value of $6.5\times10^{-10}$ as the ``best-fit'' model, this corresponds to the minimum
of the red curve of Figure~\ref{fig:chisq}.
The comparison between the observed and calculated \nnhp\:(1--0) spectral profiles is shown in 
Figure~\ref{fig:n2h+_10_best}.
The quality of the fit is remarkable: the double-peaked profile of all the hyperfine 
components is reproduced with high accuracy, including that of the weakest 
$F_1,F = (1,0)-(1,1)$ line which exhibits only a minor asymmetry due to the low optical 
opacity.
The \nnhp\ column density averaged over the IRAM 30\,m main beam is 
$N = 4.06\times10^{13}$\,cm$^{-2}$.

\begin{figure*}[t]
  \centering
  \includegraphics[angle=0,width=15cm]{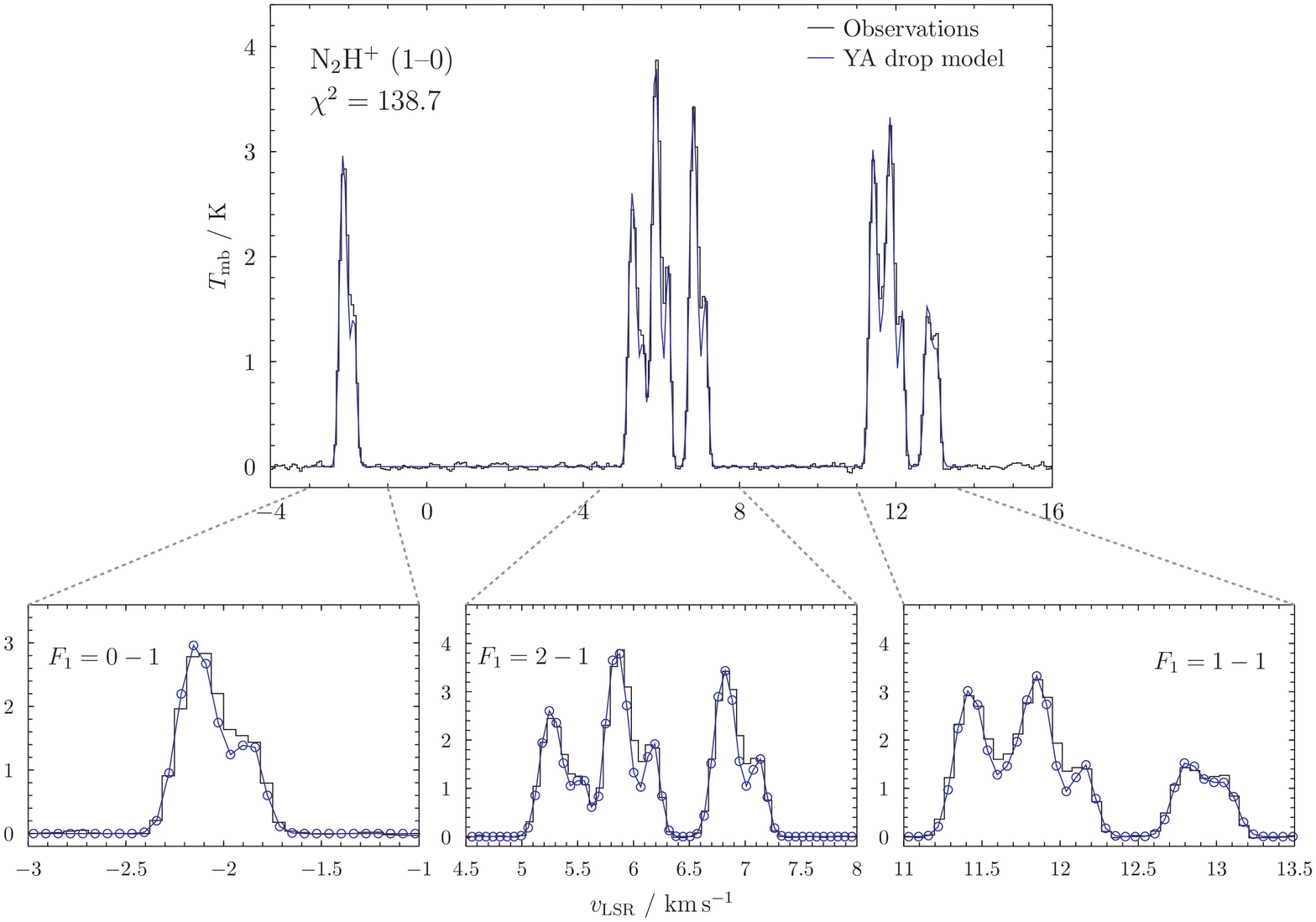}
  \caption{Observed \nnhp\:(1--0) spectrum in L1544 modelled with non-LTE code MOLLIE.
           The black histograms show our observation data.
           The blue curve shows the model spectrum computed using the ``altered'' dark 
           cloud model and the \nnhp\ radial abundance profile of \citet{Aikawa-IAU11-chem}.} 
  \label{fig:n2h+_10_2nd}
\end{figure*}

Figure~\ref{fig:n2h+_10_2nd} illustrates the optimal calculated \nnhp\:(1--0) spectrum 
adopting the centrally decreasing \nnhp\ abundance profile of \citet{Aikawa-IAU11-chem},
and larger infall velocity (yellow curve in Figure~\ref{fig:chisq}).
It can be seen that this modelling is also of reasonable quality, although the line 
profiles of the partially resolved velocity doublets is less well reproduced.
For this model, the standard abundance value is $3.0\times10^{-9}$ (value of 1 in 
Figure~\ref{fig:abund}), yielding a \nnhp\ beam averaged column density of 
$N = 3.32\times10^{13}$\,cm$^{-2}$.

Clearly, the values of $N(\nnhp)$ obtained through the above described method are affected 
by uncertainties that are difficult to evaluate.
In principle, given the least-squares procedure adopted, one may derive the 
variance--covariance matrix of the system by evaluating the Jacobian of the optimised variables, 
i.e., the \nnhp\ standard abundance, the turbulent line-width, and the scaling factor applied 
to the hydro-dynamical velocity field.

Such a purely statistical procedure yields rather small errors ($< 1$\%), very likely to be 
negligible compared to the systematic uncertainties associated to the choice of the 
spherically-symmetric hydro-dynamical model --- which we expect to be an over-simplified 
description of the real core structure --- or to those produced by the inaccuracies of the 
collision data.
To estimate the latter point, we ran various radiative transfer models, using artificially 
altered hyperfine rate coefficients by factors of~2 to~5. 
The resulting optimal \nnhp\ abundance showed only moderate changes (10-15\%, 20\% in the most 
extreme case), but the fits were systematically poorer ($\chi^2 = 200-300$) with the calculated 
spectral profile being increasingly unable to reproduce the observed infall asymmetry.

On the other hand, we have shown that one may adopt a completely different choice of \nnhp\ 
radial abundance profile still obtaining an acceptable fit (compare Figure~\ref{fig:n2h+_10_best} 
and ~\ref{fig:n2h+_10_2nd}), and this ultimately provides a mean to do an approximate evaluation 
of the $N(\nnhp)$ error bar. 
The column densities derived using different abundance profiles differ by 
$0.74\times10^{13}$\,cm$^{-2}$, i.e., 18\% of our ``best-fit'' value.
Assuming that this discrepancy represents the maximum $2\sigma$ dispersion of the actual 
$N(\nnhp)$, and adding in quadrature a conservative 10\% calibration error, we ended with 
an estimated 13\% relative uncertainty on the final results,
$N(\nnhp) = (4.1 \pm 0.5)\times 10^{13}$\,cm$^{-2}$.

\subsection{N$^{15}$NH$^+$\:(1--0) and $^{15}$NNH$^+$\:(1--0)} \label{sec:15n2h+}
\indent\indent
Given the high quality fit to the \nnhp\:(1--0) hyperfine components, we adopt the velocity 
``augmented'' L1544 model with constant molecular abundance to fit the observed spectra 
of the \isot{15}{N}-containing isotopologues.
Due to the lack of one \isot{14}{N} nucleus, the HFS of the \ninhp\:(1--0) and \innhp\:(1--0) 
lines are simpler than that of the parent species, i.e.\ they consist of triplets of hyperfine 
components.
In addition, owing to the different quadrupole coupling constants, the \ninhp\:(1--0) HFS is 
spread over 4.2\,MHz (14\,km\,s$^{-1}$), whereas the \ninhp\:(1--0) triplet are contained in 
a $\sim 1$\,MHz (3.3\,km\,s$^{-1}$) interval.
\begin{figure}[h!]
  \centering
  \includegraphics[angle=0,width=9cm]{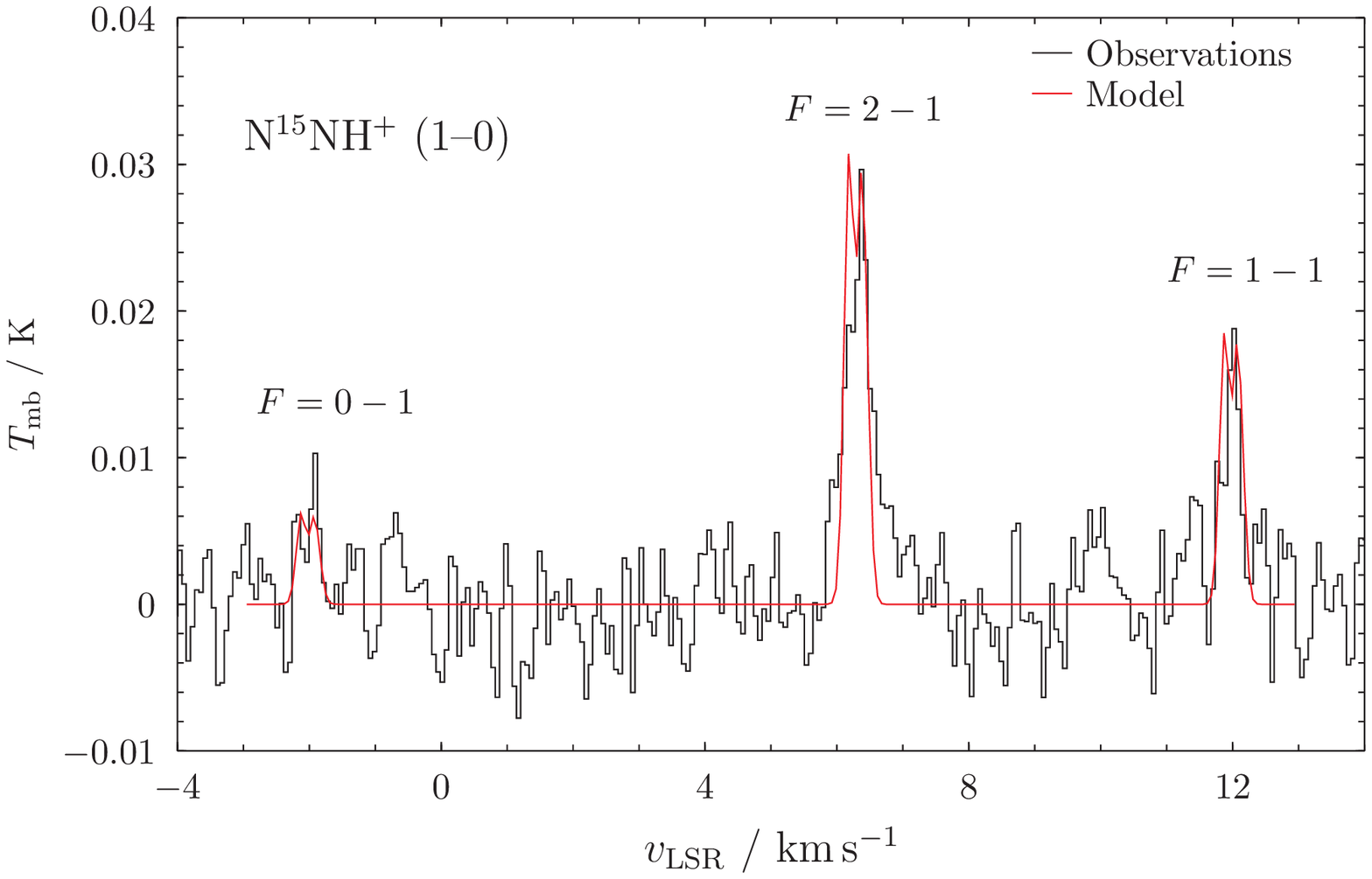}
  \caption{Observed \ninhp\:(1--0) spectrum in L1544 modelled with non-LTE code MOLLIE.
           The black histogram shows our observation data.
           The red curve shows the model spectrum computed using the best fit model 
           (see text).} 
  \label{fig:n15nh+_10_best}
\end{figure}

The best simulations of the two lines were obtained by varying independently the \ninhp\ and \innhp\ 
standard abundance until the minimum $\chi^2$ is achieved. 
The best fit dyazenilium abundances are $6.2\times 10^{-13}$ and $6.0\times 10^{-13}$, 
for \ninhp\ and \innhp, respectively.

\begin{figure}[t]
  \centering
  \includegraphics[angle=0,width=9cm]{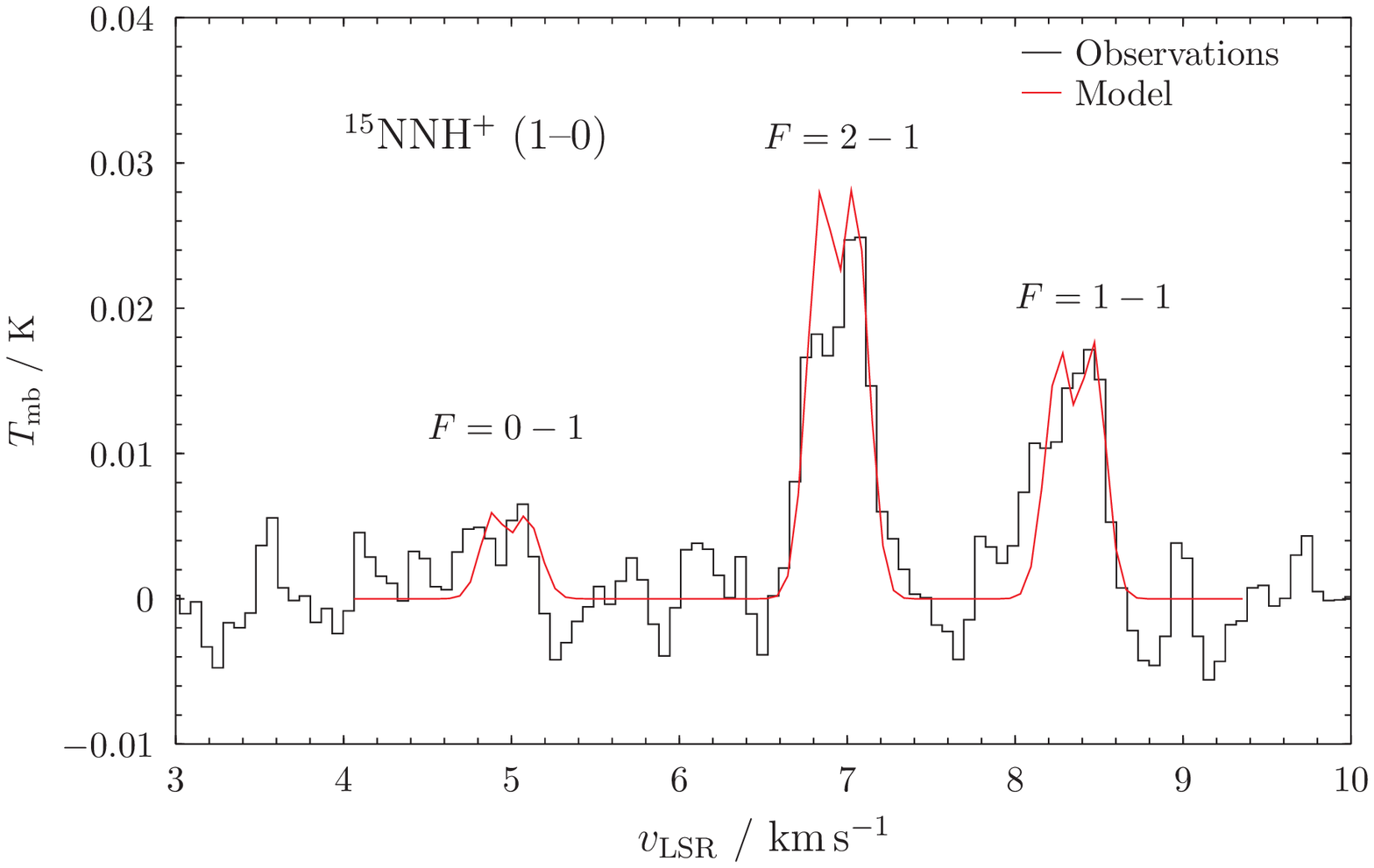}
  \caption{Observed \innhp\:(1--0) spectrum in L1544 modelled with non-LTE code MOLLIE.
           The black histogram shows our observation data.
           The red curve shows the model spectrum computed using the best fit model 
           (see text).} 
  \label{fig:15nnh+_10_best}
\end{figure}

\begin{figure}[h!]
  \centering
  \includegraphics[angle=0,width=9cm]{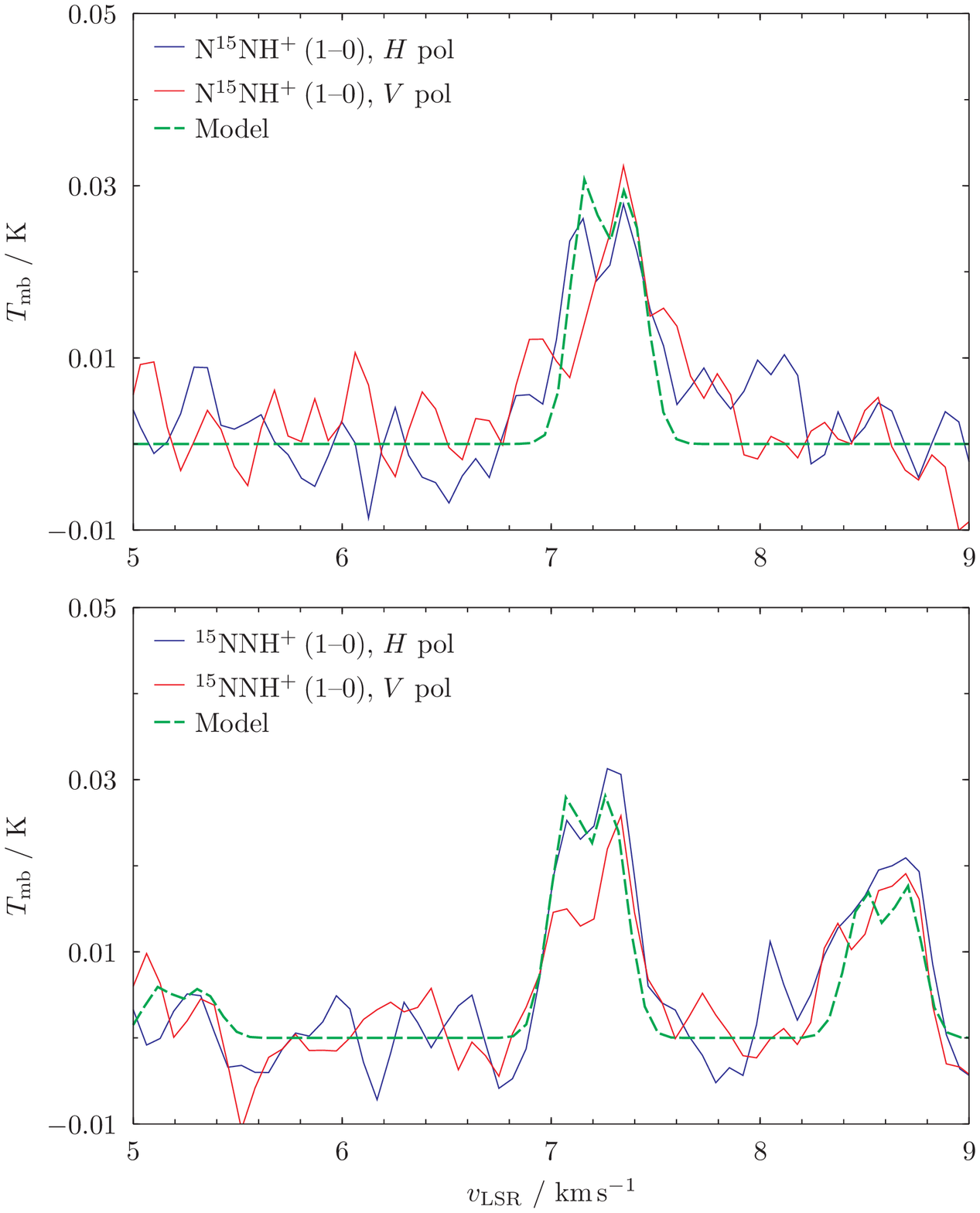}
  \caption{Detail of the central $F=2-1$ component for \ninhp\:(1--0) (upper panel) 
           and  \innhp\:(1--0) (lower panel).
           Blue and red solid curves represent the observed spectra using EMIR horizontal
           and vertical polarisation units, respectively.
           The green dashed curve shows the computed model spectra.} 
  \label{fig:var_pol}
\end{figure}

The fit results are shown in Figure~\ref{fig:n15nh+_10_best} and~\ref{fig:15nnh+_10_best}.
The model computed spectra are characterised by symmetrical line profiles; they also show a 
small but well apparent velocity doubling as expected in the case of a contracting centrally 
concentrated core.
However, the signal-to-noise ratio (SNR)  achieved by the present observations is not enough 
to reveal this feature in the L1544 spectra.
The low SNR plus the background effects are responsible for the peculiar line profile 
asymmetry exhibited by the detected components of both \isot{15}{N}-bearing dyazenilium 
variants.
A detail of the central $F=2-1$ component of the \ninhp\:(1--0) and \innhp\:(1--0) spectra
is presented in Figure~\ref{fig:var_pol}, where the data obtained with the horizontal and 
vertical polarisation units of the EMIR receiver are plotted using different curves.
It is apparent that the ``red asymmetry'' noticeable in the observed spectra is produced by
the more disturbed signal profiles provided by the vertical polarisation unit (red curves in
Figure~\ref{fig:var_pol}).

Compared to the \nnhp\:(1--0), observations of the \isot{15}{N} dyazenilium variants are 
considerably noisier and optimum molecular abundances are to be sought over broad $\chi^2$ 
minimums. 
The estimated column densities are thus affected by comparable larger error. 
Our optimisation procedure showed that the uncertainty of the \isot{15}{N}-species abundance 
determination is at most $\sim 10$\%. 
This adds in quadrature together with the other error contributions considered previously 
for the modelling of the normal isotopologue yielding a final estimate of a 17\% relative 
uncertainty of the \ninhp\ and \innhp\ column densities.

\section{Discussion} \label{sec:disc}

\subsection{Dyazenilium column densities} \label{sec:colden}
\indent\indent
Table~\ref{tab:res-mod} shows the dyazenilium column densities determined toward L1544 by 
our radiative transfer modelling and the resulting \isot{14}{N}/\isot{15}{N} ratios.
Within the estimated error bar, the abundances of \ninhp\ and \innhp\ are coincident:
$N(\ninhp)/N(\innhp) = 1.1\pm 0.3$.
The newly derived value of the \isot{14}{N}/\isot{15}{N} ratio for dyazenilium is 
$\sim 1000\pm200$, higher than the value of $446\pm 71$ previously evaluated adopting 
the LTE approximation \citep{Bizz-AA10-L1544}.
\begin{table}[tbh]
  \caption[]{Column densities determined for dyazenilium isotopologues toward L1544 and 
             derived \isot{14}{N}/\isot{15}{N} abundance ratios.}
  \label{tab:res-mod}
  \centering 
  \begin{tabular}{l c c}
    \hline\hline \noalign{\smallskip}
    Line   &  $N$ / cm$^{-2}$ & \isot{14}{N}/\isot{15}{N} \\[0.5ex]
    \hline \noalign{\smallskip}
    \nnhp   &  $(4.1 \pm 0.5)\times 10^{13}$   & $-$ \\ 
    \ninhp  &  $(3.9 \pm 0.7)\times 10^{10}$   & $1050 \pm 220$  \\
    \innhp  &  $(3.7 \pm 0.6)\times 10^{10}$   & $1110 \pm 240$  \\
    \hline
  \end{tabular}
\end{table}

In our previous letter we analysed the \ninhp\:(1--0) line emission in LTE assuming constant 
excitation temperature, $T_\mrm{ex} = 5$\,K and optical thin emission.
The obtained $N(\ninhp)$ value was $(4.1\pm 0.5)\times 10^{10}$\,cm$^{-2}$, which compares 
favourably with that derived by the present full radiative transfer modelling.
In fact, due to the associated low optical opacity, photons emitted by \isot{15}{N}-species 
can easily escape and the emerging flux is not much affected by the source structure and 
dynamics. 
LTE treatment is thus expected to yield reliable results provided that the assumed 
$T_\mrm{ex}$ is a good approximation of the average core gas temperature.

On the other hand, the \nnhp\ column density derived here is twice as large as the literature 
value of $(1.8 \pm 0.2)\times 10^{13}$\,cm$^{-2}$ obtained by \citet{Crapsi-ApJ05-L1544} 
through a LTE approach.
They derived the total optical depth, $\tau$, and $T_\mrm{ex}$ directly from the fitting 
of the $J=1-0$ hyperfine spectrum and assumed constant excitation temperature for all the
quadrupole components. 
A similar result had also been obtained previously by \citet{Caselli-ApJ02-L1544i} employing 
the integrated intensity of the ``weak'' and moderately thick $F_1,F= (1,0-1,1)$ component to 
determine the total column density using the optical thin approximation.
Both these treatment are likely to be affected by sizeable inaccuracies.
The latter method may underestimate the actual column density by a factor of 
$\sim \tau/[1-\exp(-\tau_{1,0-1,1})]$ 
($\lesssim 2$, see Appendix in \citealt{Caselli-ApJ02-L1544i}); similarly, the above estimate 
of the total optical depth is significantly uncertain as it does not take into account the 
possible presence of excitation anomalies.

Indeed, as pointed out by \citet{Daniel-ApJ06-N2H+}, for the typical gas condition prevailing 
in dark clouds, the LTE approximation is inadequate to assess molecular abundances from observed 
emission lines; in particular, the assumption of constant excitation temperatures among the 
hyperfine components of a given rotational transition fails for high opacities.
It follows that non-local computation of the radiative transfer is required to reproduce the 
intensities of the \nnhp\:(1--0) hyperfine spectrum and to retrieve reliable dyazenilium 
abundance.
We are thus confident that our revised $N(\nnhp)$ value presented in Table~\ref{tab:res-mod} 
represents a robust estimate of the actual column density of \nnhp\ toward the L1544 core.

\subsection{Nitrogen fractionation in L1544} \label{sec:Nchem}
\indent\indent
The dyazenilium \isot{14}{N}/\isot{15}{N} abundance ratio derived in the present study 
is approximately twice as large as that previously inferred \citep{Bizz-AA10-L1544},
leading us to reconsider the picture of nitrogen fractionation in L1544 and the 
implications for the chemical models.

Although only a few measurements of the \isot{14}{N}/\isot{15}{N} isotopic ratio are available, 
they clearly show a chemical differentiation.  
Nitrile-bearing species (molecule carrying the \mbox{--CN} group or its isomer) have been 
found to be considerably enriched in \isot{15}{N} 
\citep{Ikeda-ApJ02-HCN,Milam-LPI12-Nfract,HilBl-Ica13-15N}, whereas ammonia derivatives show 
no \isot{15}{N} enhancements or even a substantial depletion (e.g., \isot{15}{N}H$_2$D 
in various pre-stellar cores, \citealt{Gerin-ApJ09-15NH2D}).
This chemical dichotomy has been treated in detail by \citet{HilBl-Ica13-15N}, who proposes
a distict genesis for the \isot{15}{N}-enhancement in nitrile- and amine-bearing interstellar
molecules.
Briefly, nitriles derive from atomic nitrogen, while ammonia is formed \emph{via} N$^+$, which 
in turns come from N$_2$ (\citealt{Wirst-ApJ12-Nspin}, see also Figure~3 of 
\citealt{HilBl-Ica13-15N}).
The chemical networks responsible for their \isot{15}{N} enrichment are thus well separated.

The spin-state dependent chemical model of \citet{Wirst-ApJ12-Nspin} predict that the 
\isot{15}{N}-enrichment of ammonia is highly sensitive to the H$_2$ 
\emph{ortho}-to-\emph{para} (OPR) ratio, while the fractionation evolution of nitriles 
is not significantly affected.
The production of NH$_3$ is initiated by the ion--neutral reaction
\begin{equation} \label{eq:react1}
 \mrm{N}^+ + \mrm{H}_2 \rightarrow \mrm{NH}^+ + \mrm{H} \,,
\end{equation}
whose activation energy barrier of $\sim 200$\,K can be efficiently overcome by the 
$o$-H$_2$ internal energy. 
On the other hand, ammonia fractionation gets much less efficient as the OPR decreases and 
then an increasing quantity of \isot{15}{N}$^+$ is circulated back into molecular nitrogen 
by the equilibrium
\begin{equation} \label{eq:react2}
 ^{15}\mrm{N}^+ + ^{14}\!\mrm{N}_2 \leftrightharpoons ^{14}\!\mrm{N}^+ + ^{15}\!\mrm{N}^{14}\mrm{N} \,.
\end{equation}
The time evolution of the system shows that the $o$-H$_2$ drop is paralleled by a 
substantial rise of the ammonia \isot{14}{N}/\isot{15}{N} ratio up to double the 
original elemental fraction ($\lesssim 800$), while \isot{15}{N} enhancement of nitrile 
compounds keeps increasing reaching values in the range $\approx 100-300$.

The above reasoning matches nicely with what is observed in L1544.
Low \isot{14}{N}/\isot{15}{N} ratios have been measured for HCN 
\citep[$\sim 260$,][]{HilBl-AA10-Nchem} and HNC \citep[$>27$,][]{Milam-LPI12-Nfract}; 
contrariwise, \isot{15}{N} is under-abundant in ammonia, i.e.\ [NH$_2$D]/[$^{15}$NH$_2$D] $> 700$ 
\citep{Gerin-ApJ09-15NH2D}, suggesting an age $>2\times 10^5$\,yr for the fractionated gas 
\citep{Wirst-ApJ12-Nspin}.

In this context, our result for \nnhp\ is puzzling.
The low abundances found for \isot{15}{N}-variants suggest that the fractionation behaviour
of this ion is very similar to that of ammonia.
Their formation pathways are however distinct: NH$_3$ is generated from N$^+$ through the
parent process~\eqref{eq:react1}, whereas \nnhp\ derives from the molecular nitrogen \emph{via}
the protonation reaction
\begin{equation} \label{eq:react3}
 \mrm{N}_2 + \mrm{H}_3^+ \rightarrow \mrm{N}_2\mrm{H}^+ + \mrm{H}_2 \,,
\end{equation}
which should not be much affected by the relative abundance of $o$-H$_2$, as~\eqref{eq:react1} 
does. 
One thus might expect that the dyazenilium \isot{15}{N} content simply reflects the degree of 
fractionation of N$_2$.

The molecular nitrogen \isot{14}{N}/\isot{15}{N} ratio is predicted to lie in the interval 
100--200 by the \citet{Wirst-ApJ12-Nspin} model, matching perfectly the previous calculation
of the same authors \citep{Rodg-MNRAS08-Nfrac}.
Essentially the same result had been also obtained by \citet{Gerin-ApJ09-15NH2D} using a 
gas-phase only network.
We thus conclude that the \nnhp\ fractionation cannot be explained in the framework of 
present nitrogen chemical models.

A way to reconcile our observational results with chemical modelling is to allow selective 
freeze-out of \isot{15}{N} is some molecular form --- possibly \isot{15}{N}\isot{14}{N} --- 
on the surface of dust grains, something that needs to be tested in future models inclusive 
of \isot{15}{N}-bearing species and surface chemistry, as well as with laboratory work.

\section{Conclusion} \label{sec:conc}
\indent\indent
In this article we have reported on the detection of \innhp\ in L1544 and we have also 
presented a full non-LTE radiative transfer modelling of dyazenilium $J=1-0$ emission 
in this starless core.
Our main findings are summarised below:
\begin{enumerate}
 \item 
 The optically thick \nnhp\:(1--0) spectrum has been reproduced with a high degree of 
 accuracy using a slowly contracting Bonor-Ebert sphere to describe the cloud core. 
 The best match between observed and modelled spectrum is obtained using a constant 
 molecular abundance throughout the source.
 The skew double peak profile of the various hyperfine lines was correctly predicted 
 by this simple model; no evidence of multiple velocity components was found.
 \item 
 Our revised estimate of the \nnhp\ column density in L1544 is 
 $\sim 4\times 10^{13}$\,cm$^{-2}$, about two times higher than those determined in 
 previous investigation using simpler approaches, e.g.\ 
 LTE: $1.8\times 10^{13}$\,cm$^{-2}$, LVG: $2.7\times 10^{13}$\,cm$^{-2}$  
 \citep{Crapsi-ApJ05-L1544}.
 \item 
 \ninhp\ and \innhp\ spectra were modelled using the same method and yielded column 
 densities that agree well with our previous LTE estimates 
 ($\sim 4\times 10^{10}$\,cm$^{-2}$, \citealt{Bizz-AA10-L1544}).
 The abundance ratio between the two isotopologues is $N(\ninhp)/N(\innhp) = 1.1\pm 0.3$,
 coincident with the value of 1.25 tentatively determined by \citet{Linke-ApJ83-15N2H+} in
 DR~21\,(OH) interstellar cloud.
 The \ninhp\ enhancement predicted by \citet{Rodg-MNRAS04-Nfrac} is not observed in L1544.
 \item 
 The dyazenilium \isot{14}{N}/\isot{15}{N} ratio determined in L1544 is  $\sim 1000\pm200$. 
 This value is similar to that found for NH$_3$ \citep[$>700$][]{Gerin-ApJ09-15NH2D} and is 
 thus suggestive of a common fractionation pathway for the two molecules.
 This behaviour is not consistent with chemical models, that predict large \isot{15}{N} 
 fractionation of \nnhp. 
 We suggest that \isot{15}{N}\isot{14}{N}, the precursor for \isot{15}{N}-bearing \nnhp\ 
 molecular ions, is significantly depleted in the gas phase.
 \item 
 A set of hyperfine rate coefficients for \nnhp/H$_2$, \innhp/H$_2$ and \innhp/H$_2$ 
 collisions has been obtained from the \nnhp/He system close-coupling calculations of 
 \citet{Daniel-JCP04-N2H+,Daniel-MNRAS05-N2H+} and using a transition-dependent scaling 
 relation derived from HCO$^+$/He/H$_2$ studies.
 \end{enumerate}

\begin{acknowledgement}
The authors wish to thank the anonymous referee and the editor for the meticulous reading of 
the manuscript and the useful suggestions.
We are indebted to Eric Keto for his help with the MOLLIE code, Yuri Aikawa for supplying the 
molecular abundance profiles, Steve Charnley and Eva Wirstrom who provided the \nnhp\ 
fractionation curves calculated with their models.
We are also grateful to the IRAM $30\mut{m}$ staff for their support during the observations.
L.B. and E.L. gratefully acknowledge support from the Science and Technology Foundation 
(FCT, Portugal) through the Fellowships SFRH/BPD/62966/2009 and SFRH/BPD/71278/2010\@.
L.B. also acknowledges travel support to Pico Veleta from TNA Radio Net project funded 
by the European Commission within the FP7 Programme.
\end{acknowledgement}

\bibliographystyle{aa}
\bibliography{jsc-astro,%
              astroch,%
              poly-ions,%
              molphys,%
              misc}

\appendix

\section{Hyperfine rate coefficients for diazenylium} \label{sec:HypRates-gen}
\subsection{The general case}
\indent\indent
Hyperfine de-excitation rate coefficients for \isot{15}{N}-containing dyazenilium can be 
derived using the scattering calculations performed by \citet{Daniel-JCP04-N2H+,Daniel-MNRAS05-N2H+} 
for the parent species which has two \isot{14}{N} quadrupolar nuclei.
Using a CCSD(T) theoretically calculated potential energy surface for the \nnhp--He system, these
authors derived thermal averages of the opacity factor tensor elements $P^K_{jj'}$ up to $j = 6$
 and fitted them to an analytical form, whose coefficients were presented in a table for the ease 
of rapid evaluation of the collisional rates at various temperatures.\footnote{%
  In this appendix the lower-case symbol $j$ is used for the quantum number associated to the 
  molecule end-over-end rotation. 
  This follows the convention used in \citet{Daniel-JCP04-N2H+,Daniel-MNRAS05-N2H+} papers, in 
  which the upper-case letters are reserved to the angular momenta of the whole collisional 
  system (molecule plus atom).
}
Unfortunately, some of the equations given in \citet{Daniel-MNRAS05-N2H+} are affected by 
typographic errors, thus preventing to recover the correct values of the collisional rate 
coefficients from the given data.
With the aim of emending such inconsistencies, here we fully reproduce the derivation of the 
hyperfine rate coefficients from the data of that paper and report all the relevant equations 
in their correct form.
The reader is referred to the the original papers (and references therein) for the explanation of 
the quantities and symbols not explicitly defined here.

A convenient starting point is the expression of the hyperfine de-excitation cross-section 
\citep[see Eq.~(15) of][]{Daniel-JCP04-N2H+}:
\begin{multline}\label{eq:xsect}
  \sigma_{jF_1F\rightarrow j'F'_1F'} = \frac{\pi}{k_j^2}[F_1\,F'_1\,F'] \\
          \times \sum_K \wsxj{F_1}{F'_1}{K}{F'}{F}{I_2}^2 \wsxj{j}{j'}{K}{F'_1}{F_1}{I_1}^2
          P^K_{jj'} \,;
\end{multline}
here, $k_j$ is the wave-vector for the energy channel $E$, \mbox{$k_j^2 = (2\mu/\hbar^2)(E - E_j)$}; 
the terms in brace parentheses are the Wigner-$6j$ symbols, and the notation $[xy\ldots]$ is a handy 
shorthand for the product $(2x + 1)(y + 1)\ldots$\:\:.
Each $j\rightarrow j'$ transition comprises up to $[I_1I_2]^2 = 81$ components; they can be 
described by means of $[\mrm{Min}(j,j')]$ tensor elements $P^K_{jj'}$ with rank ranging from
$|j - j'|$ to $j + j'$ \citep{Daniel-JCP04-N2H+,Alex-JCP79-scatt,Alex-JCP83-scatt}.
This tensor contains the whole dynamics of the collisional system and is expressed in 
terms of the full atom--molecule spinless transition matrix elements $T^{\tilde{L}}_{j\,l;j'l'}$
\citep{Daniel-JCP04-N2H+,Corey-JPC83-scatt}:
\begin{multline}\label{eq:PKjj}
  P^K_{jj'} = [K] \sum_{ll'}\sum_{\tilde{L}\tilde{L}'} (-1)^{-\tilde{L}-\tilde{L}'} 
              [\tilde{L}\,\tilde{L}'] \wsxj{j}{j'}{K}{l'}{l}{\tilde{L}} 
                                      \wsxj{j}{j'}{K}{l'}{l}{\tilde{L}'}      \\
              \times T^{\tilde{L}'}_{j\,l;j'l'}{}^\ast \: T^{\tilde{L}}_{j\,l;j'l'} \,.
\end{multline}
At a given temperature, hyperfine de-excitation rates $R_{jF_1F \rightarrow j'F'_1F'}$ are 
then obtained by the convolution of the corresponding cross-sections with the Boltzmann--Maxwell 
distribution of kinetic energies:
\begin{multline}\label{eq:avg-Rh}
  R_{jF_1F \rightarrow j'F'_1F'}(T) = 
    \sqrt{\frac{8}{\mu\pi}} (k_B T)^{-3/2} \\
    \times \int_{E_{j'}}^{\infty} \sigma_{jF_1F\rightarrow j'F'_1F'}(E)\,(E - E_j)
                         \exp\left(-\frac{E - E_j}{k_B T}\right) \mrm{d}E \,,
\end{multline}
where the safe approximation $E_{j'}\approx E_{j'F'_1F'}$ has been used for the lower energy in 
the integration limits.

By substituting Eq.~\eqref{eq:xsect} for the cross sections into the integral, the expression 
of the hyperfine rate coefficients can be recast as
\begin{multline}\label{eq:Rates}
  R_{jF_1F \rightarrow j'F'_1F'} = 
    [F_1\,F'_1\,F'] \sum_K \wsxj{F_1}{F'_1}{K}{F'}{F}{I_2}^2 \wsxj{j}{j'}{K}{F'_1}{F_1}{I_1}^2 \\
    \times \left\langle \frac{\pi}{k_j^2} P^K_{jj'} \right\rangle_T 
    \mrm{e}^{E_j/k_B T} \,.       
\end{multline}
Here, the so-called Maxwellian averaged opacity factors are introduced.
Following \citet{Daniel-MNRAS05-N2H+}, they have the form
\begin{equation}\label{eq:avg-PK}
  \left\langle \frac{\pi}{k_j^2} P^K_{jj'} \right\rangle_T = 
    \sqrt{\frac{8}{\mu\pi}} (k_B T)^{-3/2} \frac{\hbar^2}{2\mu} 
    \int_{E_{j'}}^{\infty} P^K_{jj'}(E)\,\mrm{e}^{-E/k_B T} \mrm{d}E \,.
\end{equation}
Note that the constant exponential factor $\mrm{e}^{E_j/k_B T}$ is removed from the thermal
average \eqref{eq:avg-Rh} and thus appears explicitly in the definition of the collisional
rate \eqref{eq:Rates}.
 
\citet{Daniel-MNRAS05-N2H+} calculated the Maxwellian averaged opacity factors 
$\langle\pi/k_j^2 P^K_{jj'}\rangle_T$ and then fitted them to a analytical form  similar to that 
used by \citet{Grosj-AA03-H2O} \citep[see also\ ][]{Balak-ApJ99-H2,Duber-AA02-H2O}, i.e.\
\begin{equation}\label{eq:OpF-af}
  \log_{10} \left\langle \frac{\pi}{k_j^2}  P^K_{jj'} \right\rangle_T = 
      \sum_{n = 0}^{N} a^{(K,n)}_{j\leftarrow j'} x^{n - 1} \,,
\end{equation}
where $x=T^{1/3}$.
This equation is to be regarded as the formally correct expression of Eq.~(7) in 
\citet{Daniel-MNRAS05-N2H+}, and the coefficients $a^{(K,n)}_{j\leftarrow j'}$ are those 
presented in Tables~3, 4, and~5 of that paper. 
However, the reader has to considered that, for consistency, the correct header of the first 
column is $j\leftarrow j'$, and the arrows in all its entries must be reversed.

Given a temperature $T$, one can thus derive the corresponding Maxwellian averaged opacity 
factors from Eq.~\eqref{eq:OpF-af}, and then substitute back in Eq.~\eqref{eq:Rates} to obtain 
the \nnhp\ hyperfine de-excitation rates.
If needed, the corresponding excitation rates can be obtained either using Eq.~\eqref{eq:Rates} 
with the appropriate quantum number sets (the $P^K_{jj'}$ tensor is symmetrical 
by definition \eqref{eq:PKjj}), or from de-excitation rates through the detailed balance 
relation:
\begin{equation}\label{eq:detbal}
   R_{jF_1F \leftarrow j'F'_1F'} =  \frac{[F]}{[F']} R_{jF_1F \rightarrow j'F'_1F'} \,
                                    \mrm{e}^{-\Delta E/k_B T} \,,
\end{equation}
where $\Delta E$ represents the energy difference between the hyperfine levels $(jF_1F)$ and 
$(j'F'_1F')$.
By summing Eq.~\eqref{eq:Rates} over all the final states $F'_1$, $F'$, and making use of 
the 6$j$-symbol orthogonality relation, one can derive the equality:
\begin{equation}\label{eq:sum}
   \sum_{F'_1,F} R_{jF_1F \rightarrow j'F'_1F'} =  R_{j \rightarrow j'} \,,
\end{equation}
with the rotational de-excitation collisional rates defined as
\begin{equation}\label{eq:jrates}
   R_{j \rightarrow j'} = [j]^{-1} \sum_K \left\langle \frac{\pi}{k_j^2} P^K_{jj'} \right\rangle_T 
                          \mrm{e}^{E_j/k_B T} \,.       
\end{equation}
These values are coincident with those obtainable from the expansion coefficients reported in 
Table~2 of \citet{Daniel-MNRAS05-N2H+} once one considers that they define \emph{excitation} rates
and not ``de-excitation rates'' as is incorrectly stated in the caption.

\subsection{\isot{15}{N}-species}
\indent\indent
The \isot{15}{N}-variants of diazenylium contain a single quadrupolar nucleus, thus the coupling 
scheme is:
\begin{equation}\label{eq:1415-coup}
  \vec{j} + \vec{I} = \vec{F} \,,
\end{equation}
and each hyperfine level is labelled by the two quantum numbers $j$ and $F$.
The corresponding de-excitation rates, can be derived from Maxwellian averaged opacity factors 
$\langle\pi/k_j^2 P^K_{jj'}\rangle_T$ by adapting Eqs.~\eqref{eq:Rates} to the single spin case.
This is done simply by summing the right hand side of the expression over all final states 
$F'$, and again, making use of the orthogonality relation of the $6j$-symbols.
At this point, $F_1$ become the new $F$, and it results:
\begin{equation}\label{eq:R1spin}
  R_{jF \rightarrow j'F'} = [F'] \sum_K \wsxj{j}{j'}{K}{F'}{F}{I}^2
       \left\langle \frac{\pi}{k_j^2}  P^K_{jj'} \right\rangle_T \mrm{e}^{E_j/k_B T} \,.
\end{equation}
The desired rates $R_{jF \rightarrow j'F'}$ at a given $T$ are thus obtained by substituting in 
the above equality the Maxwellian average opacity factors $\langle\pi/k_j^2 P^K_{jj'}\rangle_T$ 
derived from the coefficients of \citet{Daniel-MNRAS05-N2H+} through Eq.~\eqref{eq:OpF-af}.

\subsection{Scaling of hyperfine rates}
\indent\indent
All the above data deal with the the dyazenilium/He system; to evaluate the appropriate 
rates for H$_2$ collisions we had thus to choose a reliable scaling relation.
A widely used approximation is to consider identical cross sections for these two colliding 
systems, and then apply a scaling factor of 1.37 to the rate coefficients in order to correct 
for the difference in reduced mass \citep{Lique-AA08-SiS,Schoi-AA05-LAMDA}.
However, it has been pointed out by many authors 
\citep[see, e.g.][]{Daniel-ApJ06-N2H+,Pagani-AA12-depl} that this correction is not appropriate 
for molecular ions, since the in electrostatic interaction with H$_2$ is markedly different 
from that with He.
This is indeed the case for HCO$^+$: the comparison between the HCO$^+$/He rotational rates 
calculated by \citet{Buffa-MNRAS09-HCO+} and those for the HCO$^+$/H$_2$ system 
\citep{Flower-MNRAS99-HCO+} shows that these latter are larger by a factor of about 2--5,
depending on the rotational quantum number $j$ and the temperature.

On the other hand, the HCO$^+$/He rates are practically indistinguishable from those 
of \nnhp/He \citep[hyperfine free,][]{Daniel-MNRAS05-N2H+}, due to the similarity in the 
ion electronic structure.
This suggests that accurate values of the hyperfine coefficients for the dyazenilium/H$_2$ 
system can be obtained at any given temperature as following
\begin{equation} \label{eq:scaling}
 R_{jH\rightarrow j'H'}^{(\mrm{dyaz}/\mrm{H}_2),\,\mrm{\it scaled}} = 
   R_{jH\rightarrow j'H'}^{(\mrm{dyaz}/\mrm{He})} \times
   \frac{R_{j\rightarrow j'}^{(\mrm{HCO}^+/\mrm{H}_2)}}{R_{j\rightarrow j'}^{(\mrm{HCO}^+/\mrm{He})}} \,,
\end{equation}
where $H$ represents the set of spin quantum numbers used to label a given hyperfine
level.
We derived $J\rightarrow J'$ transition dependent scaling factors from the available 
HCO$^+$ rotational data \citep{Flower-MNRAS99-HCO+,Buffa-MNRAS09-HCO+} and then applied them
to the hyperfine rates of the \nnhp/He, \ninhp/He, and \innhp/He systems calculated from the data 
of  \citet{Daniel-MNRAS05-N2H+}.
At any instance, we used Eq.~\eqref{eq:scaling} to evaluate scaled de-excitation hyperfine rates
only, whereas the corresponding coefficients for the excitation transitions where obtained through 
the detailed balance equality
\begin{equation} \label{eq:det-bal}
 R_{JH\leftarrow J'H'}^\mrm{\it scaled} = R_{JH\rightarrow J'H'}^\mrm{\it scaled}
   \frac{g(F)}{g(F')}\mrm{e}^{\Delta E/kT} \,.
\end{equation}

It should be noted that to use Eq.~\eqref{eq:scaling} one also need the elastic $\Delta J = 0$ 
rates for both hyperfine-free data sets. Such coefficients are required to scale properly the 
hyperfine rates for transitions among various $F_1$, $F$ sublevels of a given $J$.
Elastic rates for HCO$^+$/H$_2$ collisions from \citet{Flower-MNRAS99-HCO+} calculations are 
available in \textsc{basecol}\footnote{\texttt{http://basecol.obspm.fr/}} database 
\citep{Dubernet-submit}. 
Conversely, elastic coefficients were not provided for the HCO$^+$/He system in the 
\citet{Buffa-MNRAS09-HCO+} paper; we thus evaluated these rates from a set of $\Delta J = 0$ 
cross-sections provided by the author.

\section{Rate coefficient tables} \label{sec:HypRates-tab}
\indent\indent
We have calculated a full set of hyperfine de-excitation rates for the \nnhp/H$_2$, \ninhp/H$_2$, 
and \innhp/H$_2$ collision systems in the temperature range 5--50\,K.
Since these data are not included in the collisional databases, we make them available in 
electronic form at the CDS.
Tables~\ref{tab:rates1}--\ref{tab:rates3} contain the hyperfine de-excitation collisional 
rates for \nnhp/H$_2$, \ninhp/H$_2$, and \innhp/H$_2$, respectively.
Elastic rates ($\Delta J = 0, \Delta F = 0$) are also included in the tables.
The energy levels used for the calculation are reported in Table~\ref{tab:lev1}--\ref{tab:lev3}.
Energy values for \ninhp\ and \innhp\ were taken from \citet{Bizz-AA09-N2H+}, whereas for the 
parent species they were derived through a spectral fitting of the hyperfine frequencies of
\citet{Caselli-ApJ95-N2H+} and the submillimetre transitions of \citet{Amano-JMS05-N2H+}.

In this section we reproduce an excerpt of each table (the first five lines) with the 
appropriate column headings.

\begin{table}[tbh]
  \caption[]{N$_2$H$^+$/H$_2$ hyperfine rates (cm$^{3}$\,s$^{-1}$) for temperatures ranging 
             from 5\,K to 50\,K\@.}
  \label{tab:rates1}
  \begin{tabular}{rrr cccc}
    \hline\hline \noalign{\smallskip}
    $n$ & $i$ & $f$ & 5 & 10 & \ldots & 50 \\[0.5ex]
    \hline \noalign{\smallskip}
     1  &  1  &  1  & 5.480e-09 & 6.740e-09 & \ldots & 8.275e-09 \\
     2  &  2  &  2  & 5.480e-09 & 6.740e-09 & \ldots & 8.275e-09 \\
     3  &  3  &  3  & 5.480e-09 & 6.740e-09 & \ldots & 8.275e-09 \\
     4  &  4  &  1  & 0.000e+00 & 0.000e+00 & \ldots & 0.000e+00 \\
     5  &  4  &  2  & 3.107e-10 & 2.706e-10 & \ldots & 1.906e-10 \\
    $\cdots$ & $\cdots$ & $\cdots$ & $\cdots$ & $\cdots$ & $\cdots$ & $\cdots$ \\
    \hline
  \end{tabular}
  \\[1ex]
  \textbf{Notes.} $n$ is the transition number; $i$ and $f$ are the numerical labels of the 
  initial and final levels as shown in Table~\ref{tab:lev1}; the other column labels indicate 
  the absolute temperature in K.
\end{table}

\begin{table}[tbh]
  \caption[]{N$^{15}$NH$^+$/H$_2$ hyperfine rates (cm$^{3}$\,s$^{-1}$) for temperatures ranging 
             from 5\,K to 50\,K\@.}
  \label{tab:rates2}
  \begin{tabular}{rrr cccc}
    \hline\hline \noalign{\smallskip}
    $n$ & $i$ & $f$ & 5 & 10 & \ldots & 50 \\[0.5ex]
    \hline \noalign{\smallskip}
     1  &  1  &  1  & 5.480e-09 & 6.740e-09 & \ldots & 8.275e-09 \\
     2  &  2  &  1  & 3.048e-10 & 2.680e-10 & \ldots & 1.902e-10 \\
     3  &  2  &  2  & 5.552e-09 & 6.406e-09 & \ldots & 7.797e-09 \\
     4  &  3  &  1  & 3.048e-10 & 2.680e-10 & \ldots & 1.902e-10 \\
     5  &  3  &  2  & 3.837e-10 & 3.281e-10 & \ldots & 1.735e-10 \\
    $\cdots$ & $\cdots$ & $\cdots$ & $\cdots$ & $\cdots$ & $\cdots$ & $\cdots$ \\
    \hline
  \end{tabular}
  \\[1ex]
  \textbf{Notes.} $n$ is the transition number; $i$ and $f$ are the numerical labels of the 
  initial and final levels as shown in Table~\ref{tab:lev2}; the other column labels indicate 
  the absolute temperature in K.
\end{table}

\begin{table}[tbh]
  \caption[]{$^{15}$NNH$^+$/H$_2$ hyperfine rates (cm$^{3}$\,s$^{-1}$) for temperatures ranging 
             from 5\,K to 50\,K\@.}
  \label{tab:rates3}
  \begin{tabular}{rrr cccc}
    \hline\hline \noalign{\smallskip}
    $n$ & $i$ & $f$ & 5 & 10 & \ldots & 50 \\[0.5ex]
    \hline \noalign{\smallskip}
     1  &  1  &  1  & 5.480e-09 & 6.740e-09 & \ldots & 7.175e-09 \\
     2  &  2  &  1  & 3.022e-10 & 2.668e-10 & \ldots & 2.497e-10 \\
     3  &  2  &  2  & 5.502e-09 & 6.375e-09 & \ldots & 6.697e-09 \\
     4  &  3  &  1  & 3.022e-10 & 2.668e-10 & \ldots & 2.497e-10 \\
     5  &  3  &  2  & 3.803e-10 & 3.266e-10 & \ldots & 2.833e-10 \\
    $\cdots$ & $\cdots$ & $\cdots$ & $\cdots$ & $\cdots$ & $\cdots$ & $\cdots$ \\
    \hline
  \end{tabular}
  \\[1ex]
  \textbf{Notes.} $n$ is the transition number; $i$ and $f$ are the numerical labels of the 
  initial and final levels as shown in Table~\ref{tab:lev3}; the other column labels indicate 
  the absolute temperature in K.
\end{table}

\begin{table}[tbh]
  \caption[]{N$_2$H$^+$ hyperfine level energies (MHz).}
  \label{tab:lev1}
  \begin{tabular}{r c ccc}
    \hline\hline \noalign{\smallskip}
    $n$ &   Energy    & $J$ & $F_1$ & $F$ \\[0.5ex]
    \hline \noalign{\smallskip}
     1  &      0.0    &  0  &   1   &  0  \\
     2  &      0.0    &  0  &   1   &  1  \\
     3  &      0.0    &  0  &   1   &  2  \\
     4  &  93171.6198 &  1  &   1   &  0  \\
     5  &  93171.9173 &  1  &   1   &  2  \\
     $\cdots$ & $\cdots$ & $\cdots$ & $\cdots$ & $\cdots$ \\
    \hline
  \end{tabular}
\end{table}

\begin{table}[tbh]
  \caption[]{N$^{15}$NH$^+$ hyperfine level energies (MHz).}
  \label{tab:lev2}
  \begin{tabular}{r c cc}
    \hline\hline \noalign{\smallskip}
    $n$ &   Energy    & $J$ &  $F$ \\[0.5ex]
    \hline \noalign{\smallskip}
     1  &       0.0    &  0  &  1 \\
     2  &   91204.2602 &  1  &  1 \\
     3  &   91205.9909 &  1  &  2 \\
     4  &   91208.5160 &  1  &  0 \\
     5  &  273613.6252 &  2  &  2 \\
     $\cdots$ & $\cdots$ & $\cdots$ & $\cdots$ \\
    \hline
  \end{tabular}
\end{table}

\begin{table}[tbh]
  \caption[]{$^{15}$NNH$^+$ hyperfine level energies (MHz).}
  \label{tab:lev3}
  \begin{tabular}{r c cc}
    \hline\hline \noalign{\smallskip}
    $n$ &   Energy    & $J$ &  $F$ \\[0.5ex]
    \hline \noalign{\smallskip}
     1  &       0.0    &  0  &  1 \\
     2  &   90263.4870 &  1  &  1 \\
     3  &   90263.9120 &  1  &  2 \\
     4  &   90264.4973 &  1  &  0 \\
     5  &  270789.1759 &  2  &  2 \\
     $\cdots$ & $\cdots$ & $\cdots$ & $\cdots$ \\
    \hline
  \end{tabular}
\end{table}

\end{document}